\documentclass[reprint,superscriptaddress,amsmath,amssymb,twocolumn,prl]{revtex4-2}

\usepackage{amsmath}
\usepackage{mathtools}
\usepackage{amssymb}
\usepackage{graphicx}
\usepackage{bm}
\usepackage[colorlinks,linkcolor=blue,citecolor=blue,urlcolor=blue,breaklinks=true]{hyperref}
\usepackage{booktabs}
\usepackage{multirow}
\usepackage[dvipsnames]{xcolor}
\usepackage{ragged2e}
\usepackage{orcidlink}
\usepackage[T1]{fontenc}
\usepackage{mathrsfs}

\newcommand{\Tr}{\operatorname{Tr}}
\newcommand{\cG}{\mathcal G}
\newcommand{\Aenergy}{\mathcal A_E}
\newcommand{\Ires}{\mathcal I_{E,R}}
\newcommand{\Cset}{\mathfrak C}
\newcommand{\figpanel}[2]{\hyperref[#1]{\ref*{#1}(#2)}}

\begin{document}

\date{\today}

\title{Finite-Particle Quantum Reduction of Thermodynamic Irreversibility}

\author{Borhan Ahmadi\orcidlink{0000-0002-2787-9321}}
\email{borhan.ahmadi@ug.edu.pl}
\address{International Centre for Theory of Quantum Technologies, University of Gda\'nsk, ul. prof. Marii Janion 4, 80-309 Gda\'nsk, Poland}

\begin{abstract}
Thermodynamics is useful because it lets us predict and control complex systems from a few accessible quantities when microscopic reconstruction is impractical. We ask whether this operational reduction can be less costly in a finite quantum system than in a matched classical one. We compare a few-boson Bose--Hubbard chain with its number-conserving classical-field counterpart under the same driving and the same coarse spatial record. The quantum system relaxes more in the observed particle distribution, yet the retained mean energy resolves more of the microscopic structure hidden by that record, leaving less information unusable and smaller entropy generation. We then close the process into a heat-engine cycle whose controls use only the measured mean energy and spatial record, not full state tomography; at matched thermal resources, the smaller entropy generation gives more work and higher efficiency. The advantage fades as the particle number grows toward the classical-field regime.
\end{abstract}

\maketitle


\textit{Introduction}.--
Thermodynamics becomes useful precisely when microscopic description, measurement, or control is impractical. Instead of tracking every microscopic detail, we retain a small set of measurable variables and ask what can still be predicted and controlled from them. This reduction should not be confused with approximating the microscopic state: a coarse description can represent the retained information exactly while deliberately leaving the rest unused. The price is that information outside the thermodynamic record cannot help the controller. This is the old tension behind irreversibility. Clausius introduced entropy to distinguish reversible and irreversible changes, while Boltzmann and Gibbs connected thermodynamics to microscopic states and their multiplicities~\cite{Clausius1865,Boltzmann1877,Gibbs1902}. The microscopic dynamics can preserve fine information even when the reduced description develops an arrow of time. The practical question is then: once we choose what the thermodynamic observer keeps, how much useful control remains possible?

Several standard ideas make this question precise. Jaynes' maximum-entropy principle gives the least biased state consistent with specified macroscopic data~\cite{Jaynes1957}. Zwanzig separated an ensemble into a part fixed by chosen observables and a part that remains unresolved~\cite{Zwanzig1960}. Robertson later built generalized canonical states directly from time-dependent thermodynamic variables~\cite{Robertson1966}. In all of these approaches, the thermodynamic state summarizes retained information; it need not approximate, or coincide with, the exact microscopic state.

The same distinction is important in isolated quantum systems. Von Neumann's macrospace construction connected coarse quantum descriptions with an \(H\)-theorem-like structure~\cite{vonNeumann1929}. Later work showed that limits on measurements can change how quantum and classical equilibration appear~\cite{ReimannEvstigneev2013}. Observational entropy made the chosen coarse description explicit, and local particle number followed by total energy was shown to give a useful thermodynamic entropy for isolated quantum dynamics~\cite{Safranek2019thermo}. A classical version and a quantum--classical correspondence were then developed~\cite{Safranek2020classical}, and the associated coarse state was linked to recoverability~\cite{Buscemi2023}. More recent work has derived second-law-like behavior for isolated quantum systems when only restricted observables are kept~\cite{Meier2025}. We therefore do not introduce a new coarse entropy or a new maximum-entropy principle.

Limited information matters only thermodynamically if it changes what can actually be done. Restricting the available Hamiltonians changes work-extraction bounds~\cite{Wilming2016}, and coarse measurements can be enough to determine the work value of incompletely known quantum sources~\cite{SafranekRosaBinder2023}. Other recent approaches make thermodynamic quantities depend explicitly on measurement precision or microscopic access~\cite{Rubino2026,PernambucoCeleri2026}, and closed-system thermodynamic laws have been derived for restricted macroscopic observables and operations~\cite{3y8f-yrp8}. Hidden athermality and work can likewise be organized by levels of access~\cite{HokkyoUeda2026}. At the same time, incomplete knowledge does not always reduce extractable work: in suitable many-copy settings, state-agnostic protocols can reach the optimal asymptotic rate~\cite{WatanabeTakagi2026}. Any thermodynamic cost of limited information must therefore be tied to the record and controls that are actually used.

Here we make that operational test in a matched quantum--classical setting. We compare a finite quantum system with a classical model of the same many-body physics, driven by the same work protocol and viewed through the same coarse record. The question is not only whether their coarse entropies differ, but whether the same restricted information supports different thermodynamic performance. We build on Ref.~\cite{ahmadi2026endogenous}, where exact microscopic dynamics preserves fine information and irreversibility is defined relative to a chosen thermodynamic record.

Our answer is yes for the system studied here. We compare a finite-particle Bose--Hubbard chain with its number-conserving classical Hamiltonian-field limit. The two models share the same coherent-state energy landscape, the same confinement protocol, and the same coarse spatial record. We use one maximum-entropy state for the spatial record alone and a second one that also keeps the mean energy. This separates the information hidden by the spatial record into a part that the energy can still resolve and a part that remains hidden even after the energy is kept. The quantum particle distribution relaxes more strongly, yet the mean energy resolves more of the hidden microscopic structure, leaving a smaller residual information gap and smaller process entropy generation. Crucially, we then close the process into a cycle whose controls after the nonequilibrium endpoint use only the retained mean energy and record probabilities, not full state tomography. At matched thermal resources, the smaller entropy generation gives more work and higher efficiency. The advantage becomes smaller as the particle number grows toward the classical-field regime.

\textit{Cycle and matched quantum--classical dynamics}.--
We first describe the physical cycle so that every state and control step is clear. Two records are used. The spatial record \(R\) keeps only the coarse particle distribution between the two halves of the lattice. The cycle controller keeps this record together with the mean energy; we denote the combined information by \((E,R)\). The exact microscopic state exists throughout the cycle, but the controller does not reconstruct the information outside the chosen record.

The sequence is \(A\rightarrow B\rightarrow C^-\rightarrow C^+\rightarrow\bar C\rightarrow D\rightarrow A\). Point \(A\) is an equilibrium state at the hot temperature \(T_h\). The branch \(A\rightarrow B\) is a reversible hot isotherm. At \(B\), a positive confinement makes occupation of the right half costly. The working medium is then isolated, and the confinement is lowered in a finite time during \(B\rightarrow C^-\). The point \(C^-\) is the actual nonequilibrium endpoint of this isolated work process.

The cold side is split into steps only to make the physics explicit. From the mean energy and record probabilities measured at \(C^-\), the controller constructs the energy-and-record maximum-entropy representative and a Hamiltonian for which this representative is Gibbs at \(T_c\). While the working medium is still isolated, the Hamiltonian is changed in the work-only switch \(C^-\rightarrow C^+\). The switch changes the Hamiltonian but leaves the microscopic state unchanged. At \(C^+\), the cold reservoir is attached once and stays attached until \(D\). First, with the controlled Hamiltonian fixed, the actual state relaxes to the energy-and-record representative; this equilibrium point is called \(\bar C\). The same cold contact is then kept while the Hamiltonian is changed quasistatically from \(\bar C\) to the cold equilibrium point \(D\). The reservoir is removed at \(D\), and a reversible isolated scaling of the Hamiltonian returns the system to \(A\).

The quantum working medium is an open Bose--Hubbard chain of \(L\) sites with exactly \(Q\) bosons (\(\hbar=1\))~\cite{Fisher1989,Jaksch1998},
\begin{align}
H_Q(\Lambda)
&=
-J\sum_{j=1}^{L-1}
\left(
a_j^\dagger a_{j+1}
+
a_{j+1}^\dagger a_j
\right)
\nonumber\\
&\quad+
\frac{U_Q}{2}
\sum_{j=1}^{L}
n_j(n_j-1)
+
\Lambda N_R ,
\end{align}
where \(J\) is the hopping amplitude, \(n_j=a_j^\dagger a_j\), and \(N_R=\sum_{j=L/2+1}^{L}n_j\) counts particles in the right half. The external parameter \(\Lambda\) controls the confinement.

The classical state is a normalized complex field \(z=(z_1,\ldots,z_L)\) with \(\sum_j|z_j|^2=1\). Its common phase has no physical effect, so the phase space is \(CP^{L-1}\). With \(x_R(z)=\sum_{j>L/2}|z_j|^2\), the Hamiltonian per particle is~\cite{Trombettoni2001,Trimborn2008}
\begin{equation}
h_\Lambda(z)
=
-J\sum_{j=1}^{L-1}
\left(
z_j^\ast z_{j+1}
+
z_{j+1}^\ast z_j
\right)
+
\frac{g}{2}
\sum_{j=1}^{L}
|z_j|^4
+
\Lambda x_R(z).
\end{equation}
Hamilton's equations give the number-conserving discrete nonlinear Schr\"odinger dynamics, \(i\dot z_j=-J(z_{j-1}+z_{j+1})+g|z_j|^2z_j+\Lambda(t)\chi_R(j)z_j\), with missing neighbors omitted and \(\chi_R(j)=1\) on the right half.

The quantum and classical Hamiltonians are matched before their dynamics are compared. For the number-conserving coherent state \(|z;Q\rangle=(\sum_jz_ja_j^\dagger)^Q|0\rangle/\sqrt{Q!}\), choosing \(U_Q=g/(Q-1)\) gives
\begin{equation}
\frac{\langle z;Q|H_Q(\Lambda)|z;Q\rangle}{Q}
=
h_\Lambda(z).
\end{equation}
Thus a classical configuration and the corresponding coherent state have the same energy per particle. Their dynamics are still different: the quantum state explores the full fixed-\(Q\) Hilbert space, while the classical field follows Hamiltonian flow on \(CP^{L-1}\). In the number-conserving phase-space description of the Bose--Hubbard model, the finite-\(Q\) corrections vanish as \(Q\) grows, leaving the classical Liouville dynamics~\cite{Trimborn2008}.

\begin{figure*}[t]
\centering
\includegraphics[width=1\textwidth]{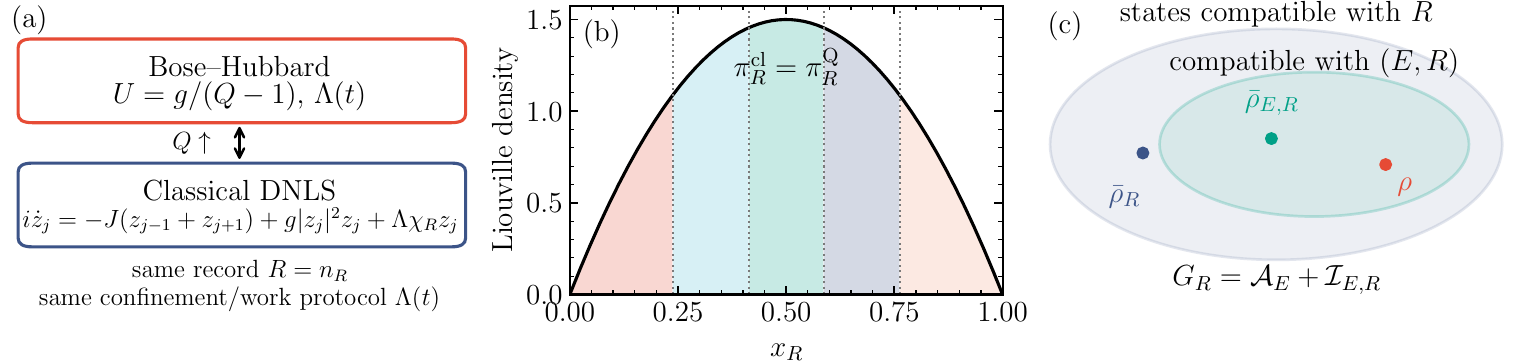}
\caption{
Matched quantum--classical construction.
(a) The finite-particle Bose--Hubbard model and its number-conserving classical-field counterpart have the same coherent-state energy landscape and undergo the same confinement protocol.
(b) The classical right-half coordinate is divided so that each region has the same invariant weight as the matching quantum occupation sector.
(c) Three states associated with the same microscopic situation. The exact state \(\rho\) contains the full information. The record-only representative \(\bar\rho_R\) keeps only the spatial probabilities. The energy-and-record representative \(\bar\rho_{E,R}\) also keeps the exact mean energy. The difference \(\bar\rho_{E,R}-\bar\rho_R\) is the refinement supplied by the energy, while \(\rho-\bar\rho_{E,R}\) remains hidden from the combined \((E,R)\) description.\justifying}
\label{fig:framework}
\end{figure*}

At \(B\), the working medium is in thermal equilibrium with the hot reservoir at confinement \(\Lambda_0\). We keep the intensive temperature scales \(\theta_h\) and \(\theta_c\) fixed through \(k_{\rm B}T_h=Q\theta_h\) and \(k_{\rm B}T_c=Q\theta_c\). The quantum preparation is \(\rho_B^{\rm Q}=e^{-H_Q(\Lambda_0)/(k_{\rm B}T_h)}/Z_B^{\rm Q}\), and the classical preparation is \(f_B^{\rm cl}(z)=e^{-h_{\Lambda_0}(z)/\theta_h}/Z_B^{\rm cl}\) with respect to the normalized invariant measure on \(CP^{L-1}\). During \(B\rightarrow C^-\), the working medium is isolated and \(\Lambda(t)\) is lowered from \(\Lambda_0\) to zero over the time \(\tau_{BC}\). Because no heat crosses the working-medium boundary, the energy transferred by changing \(\Lambda\) is work. The fixed-particle Hilbert space, coherent-state map, classical equations, temperature scaling, and many-particle limit are given in Note~\ref{supp:note-classical-limit} of the SM \cite{SuppMat}. Figure~\figpanel{fig:framework}{a} summarizes the two microscopic descriptions.

\textit{A common thermodynamic record}.--
A fair comparison needs the same coarse information on both sides. We therefore ask the same question of the quantum and classical systems: how much of the working medium is in the right half?

For a quantum state \(\rho\), the observable \(N_R\) has outcomes \(r=0,1,\ldots,Q\). Let \(P_r\) project onto the subspace with outcome \(r\), and let \(p_r=\Tr(P_r\rho)\). The probabilities \(\{p_r\}\) form the spatial record \(R\). The richer record \((E,R)\) also keeps the exact mean energy. It still contains far less information than full state tomography.

For two equal halves with \(\ell=L/2\) sites each, the number of quantum many-body states with right-half occupation \(r\) is
\begin{equation}
\Omega_r^{\rm Q}
=
\binom{r+\ell-1}{r}
\binom{Q-r+\ell-1}{Q-r},
\quad
\Omega_{\rm tot}^{\rm Q}
=
\binom{Q+L-1}{Q}.
\end{equation}
The corresponding reference weight is \(\pi_r^{\rm Q}=\Omega_r^{\rm Q}/\Omega_{\rm tot}^{\rm Q}\). It is a counting weight, not a thermal distribution.

The classical right-half fraction \(x_R\) is continuous, so we divide it into \(Q+1\) regions. Under the uniform invariant measure on \(CP^{2\ell-1}\), \(x_R\) has density \(w_\ell(x)=\Gamma(2\ell)x^{\ell-1}(1-x)^{\ell-1}/\Gamma(\ell)^2\). We choose the boundaries \(b_r\) so that each classical region \(A_r=\{z:b_r\leq x_R(z)<b_{r+1}\}\) has the same invariant weight as the matching quantum sector,
\begin{equation}
\int_{b_r}^{b_{r+1}} w_\ell(x)\,dx
=
\pi_r^{\rm Q}.
\end{equation}
Hence \(\pi_r^{\rm cl}=\mu(A_r)=\pi_r^{\rm Q}\). The partition is fixed before any nonequilibrium dynamics is generated. For \(L=4\) and \(Q=4\), the common reference is \(\pi=(5,8,9,8,5)/35\). Figure~\figpanel{fig:framework}{b} shows the corresponding partition (see Note~\ref{supp:note-matched-record} of the SM \cite{SuppMat} for details).

The two systems now share the same energy landscape, driving protocol, coarse physical record, and reference weight for every record outcome. Any remaining difference comes from the dynamics and from how the same retained information constrains the two microscopic state spaces.

\textit{Thermodynamic states from limited information}.--
Let \(\rho\) be the exact quantum state at some time. We do not replace it by a coarse state. Instead, we ask which microscopic states are compatible with the information that the thermodynamic observer keeps.

If only the probabilities \(p_r\) are retained, the compatible states are
\begin{equation}
\Cset_R(p)
=
\left\{
\sigma\geq0:
\Tr\sigma=1,\quad
\Tr(P_r\sigma)=p_r\ \forall r
\right\}.
\end{equation}
The maximum-entropy state in this set is
\begin{equation}
\bar\rho_R
=
\cG_R[\rho]
=
\sum_r
\frac{p_r}{\Omega_r^{\rm Q}}P_r,
\quad
S_R
=
-k_{\rm B}
\sum_r
p_r
\ln\!\left(
\frac{p_r}{\Omega_r^{\rm Q}}
\right).
\end{equation}
We call \(\bar\rho_R\) the record-only representative. The exact state can be written as \(\rho=\bar\rho_R+\chi_R\), where \(\chi_R=\rho-\bar\rho_R\) and \(\Tr(P_r\chi_R)=0\) for every \(r\). Thus \(\chi_R\) contains microscopic structure that the spatial record does not see. This is an exact decomposition, not a physical projection of the state.

If the exact mean energy \(E=\Tr(H\rho)\) is also retained, the compatible set becomes
\begin{equation}
\Cset_{E,R}(E,p)
=
\left\{
\sigma\in\Cset_R(p):
\Tr(H\sigma)=E
\right\}.
\end{equation}
Its maximum-entropy state is the energy-and-record representative. For an interior solution,
\begin{equation}
\bar\rho_{E,R}
=
\frac{
\exp[-\beta^\star H-\sum_r\alpha_rP_r]
}{
Z_{E,R}
}.
\end{equation}
The multipliers are chosen so that this state reproduces the same mean energy and the same \(p_r\) as \(\rho\). Away from equilibrium, \(\beta^\star\) is only the multiplier that enforces the energy constraint; it is not by itself a physical inverse temperature. We denote the entropy of \(\bar\rho_{E,R}\) by \(S_{E,R}\).

The richer record gives a second exact split, \(\rho=\bar\rho_{E,R}+\chi_{E,R}\), with \(\Tr(P_r\chi_{E,R})=0\) and \(\Tr(H\chi_{E,R})=0\). Defining \(\delta_{E|R}=\bar\rho_{E,R}-\bar\rho_R\), the two decompositions give
\begin{equation}
\chi_R
=
\delta_{E|R}
+
\chi_{E,R}.
\end{equation}
The first term is the change in the representative caused by adding the measured energy. The second is what remains unresolved even after both \(E\) and \(R\) are kept. These are trace-zero Hermitian operator differences, not states by themselves.

The three states \(\rho\), \(\bar\rho_{E,R}\), and \(\bar\rho_R\) should not be confused. Their entropies obey \(S_{\rm vN}(\rho)\leq S_{E,R}\leq S_R\). At canonical equilibrium, the exact state already maximizes entropy at its own mean energy, so \(\rho_{\rm eq}=\bar\rho_{E,R}\). In general, however, \(\rho_{\rm eq}\neq\bar\rho_R\), because the spatial record alone does not fix the energy.

The classical construction is the same. For the exact Liouville density \(f(z)\), the record-only representative \(\bar f_R\) is uniform inside each region \(A_r\), while \(\bar f_{E,R}\propto\exp[-\beta^\star h-\alpha_{R(z)}]\) also reproduces the mean energy. The corresponding fine entropy is \(S_{\rm fine}^{\rm cl}=-k_{\rm B}\int f\ln f\,d\mu\). The record-only map \(\cG_R\) is linear and is the map used in the exact finite-step coarse dynamics of Ref.~\cite{ahmadi2026endogenous}. The energy-and-record assignment is generally nonlinear because its multipliers depend on the current constraints. We therefore use it as a nested thermodynamic description, not as a replacement for \(\cG_R\) in that finite-step dynamics (see Note~\ref{supp:note-pythagorean} of the SM \cite{SuppMat} for the full derivation of the relations).

\textit{What the mean energy resolves}.--
We now turn the three-state picture into three entropy gaps. To keep a state property separate from entropy generated along a process, we introduce
\begin{equation}
\begin{aligned}
G_R
&\equiv
S_R-S_{\rm vN}
=
k_{\rm B}D(\rho\Vert\bar\rho_R),
\\
\Aenergy
&\equiv
S_R-S_{E,R}
=
k_{\rm B}D(\bar\rho_{E,R}\Vert\bar\rho_R),
\\
\Ires
&\equiv
S_{E,R}-S_{\rm vN}
=
k_{\rm B}D(\rho\Vert\bar\rho_{E,R}).
\end{aligned}
\end{equation}
Here \(\Ires\) is a single-state quantity: it is the residual information gap left after both the mean energy and the spatial record are retained. We reserve the symbol \(\Sigma\) for entropy generation associated with a process or a complete cycle. The three state gaps obey the exact identity
\begin{equation}
G_R
=
\Aenergy+\Ires.
\end{equation}
The same identity holds classically with the Kullback--Leibler divergence (see Note~\ref{supp:note-pythagorean} of the SM \cite{SuppMat} for a proof, including support and boundary cases).

The equality conditions are simple. \(G_R=0\) only when \(\rho=\bar\rho_R\), \(\Ires=0\) only when \(\rho=\bar\rho_{E,R}\), and \(\Aenergy=0\) only when the two representatives are the same. At canonical equilibrium, \(\rho_{\rm eq}=\bar\rho_{E,R}\), so \(\Ires=0\), while \(G_R=\Aenergy\) can still be nonzero.

We now apply this to the isolated process \(B\rightarrow C^-\). At \(B\), both working media are equilibrium maximum-entropy states for their own \((E,R)\) data, so \(\Ires^X(B)=0\) for \(X\in\{{\rm Q},{\rm cl}\}\). During the isolated evolution, the fine quantum and classical entropies remain constant. At \(C^-\), the exact state and the two representatives are generally different.

We define the record-relative entropy generation of this process by
\begin{equation}
\begin{aligned}
\Sigma_{BC}^X
&\equiv
S_{E,R}^X(C^-)-S_{E,R}^X(B)
\\
&=
\Ires^X(C^-)-\Ires^X(B)
=
\Ires^X(C^-)
\\
&=
\Delta G_R^X-\Delta\Aenergy^X .
\end{aligned}
\end{equation}
The subscript \(BC\) is kept for the plotted process quantity; its physical endpoint is \(C^-\). The first line defines a process entropy generation. The quantity \(\Ires(C^-)\), by contrast, is a state property of the endpoint. They are equal here because the fine entropy is conserved during the isolated process and the initial equilibrium state has \(\Ires(B)=0\). Quantum mechanically this gives \(\Sigma_{BC}^{\rm Q}=k_{\rm B}D[\rho_{C^-}^{\rm Q}\Vert\bar\rho_{E,R}^{\rm Q}(C^-)]\), with the classical Kullback--Leibler analogue. This equality does not mean that fine microscopic entropy is produced during the isolated dynamics.

The decomposition also tells us how a quantum reduction can arise. The quantum dynamics can hide less information from the spatial record, or the mean energy can resolve more of the information hidden by that record. The calculations show that the second effect is the important one.

\textit{Finite-particle quantum reduction and its mechanism}.--
We now apply the construction above to a concrete isolated work process.
We first consider \(Q=4\) bosons on \(L=4\) sites.
The quantum and classical working media are prepared at point \(B\) in their
canonical equilibrium states for the confined Hamiltonians
\(H_Q(\Lambda_0)\) and \(h_{\Lambda_0}\) defined above.
The working medium is then isolated, and during \(B\rightarrow C^-\) the
confinement is lowered from \(\Lambda_0\) to zero over the finite time
\(\tau_{BC}\). Thus the initial state is thermal, while the endpoint \(C^-\)
is generated dynamically by the same confinement protocol in the matched
quantum and classical descriptions. We first compare their natural thermal
preparations and then impose a stricter preparation in which the accessible
initial data are matched exactly.

For the representative parameters we take
\(g=1.3J\), \(\Lambda_0=20J\),
\(\theta_h=1.25J\), \(\theta_c=0.375J\), and
\(J\tau_{BC}=20\). With the matching
\(U_Q=g/(Q-1)\), these choices give
\(U_Q/J=1.3/3\), \(k_{\rm B}T_h=5J\), and
\(k_{\rm B}T_c=1.5J\).

For the natural thermal preparations, the isolated expansion gives
\(\Sigma_{BC}^{\rm Q}/k_{\rm B}=0.641610\) and
\(\Sigma_{BC}^{\rm cl}/k_{\rm B}=0.778928\pm0.003298\).
Thus, after the same finite-time confinement ramp, the retained
\((E,R)\) description assigns a smaller entropy generation to the
finite-particle quantum dynamics (see Notes~\ref{supp:note-quantum-numerics} and \ref{supp:note-classical-numerics} of the SM \cite{SuppMat} for the numerical procedures).

We next use
\(\Sigma_{BC}=\Delta G_R-\Delta\Aenergy\)
to identify the origin of this difference.
Quantum mechanically,
\(\Delta G_R^{\rm Q}/k_{\rm B}=1.737744\) and
\(\Delta\Aenergy^{\rm Q}/k_{\rm B}=1.096134\).
Classically, the corresponding values are
\(1.569713\pm0.000436\) and
\(0.790785\pm0.003180\).
Hence
\begin{equation}\nonumber
\frac{\Sigma_{BC}^{\rm cl}-\Sigma_{BC}^{\rm Q}}{k_{\rm B}}
=
\underbrace{-0.168031}_{\text{record}}
+
\underbrace{0.305350}_{\text{energy-resolved}}
=
0.137318.
\end{equation}
The two terms have opposite signs. The spatial record alone actually
favors the classical system: more information becomes hidden from \(R\)
on the quantum side. The final ordering is reversed only after the mean
energy is retained, because energy resolves a still larger part of that
record-hidden information quantum mechanically. The smaller quantum
\(\Sigma_{BC}\) therefore does not come from weaker coarse relaxation,
but from the greater resolving power of the combined \((E,R)\) record
at the nonequilibrium endpoint.

This behavior is not built into the use of mean energy. Energy adds one
scalar constraint in both descriptions, but how strongly that constraint
narrows the set of record-compatible microscopic states depends on the
Hamiltonian and on the chosen record. It may add substantial information,
very little, or none (see Note~\ref{supp:no-universal-ordering} of the
SM \cite{SuppMat} for a simple counterexample).

The natural quantum and classical canonical states belong to different
microscopic state spaces, so their accessible data at \(B\) are not
identical. We therefore repeat the comparison with a stricter initial
matching. The quantum state is left unchanged, while the classical
initial density is chosen as the maximum-entropy state having exactly
the same record probabilities \(p_r(B)\) and the same mean energy as
the quantum state. The two systems therefore enter the isolated
\(B\rightarrow C^-\) stroke with the same thermodynamic information,
while their microscopic dynamics remain quantum and classical,
respectively.

\begin{figure}[t]
\centering
\includegraphics[width=1\linewidth]{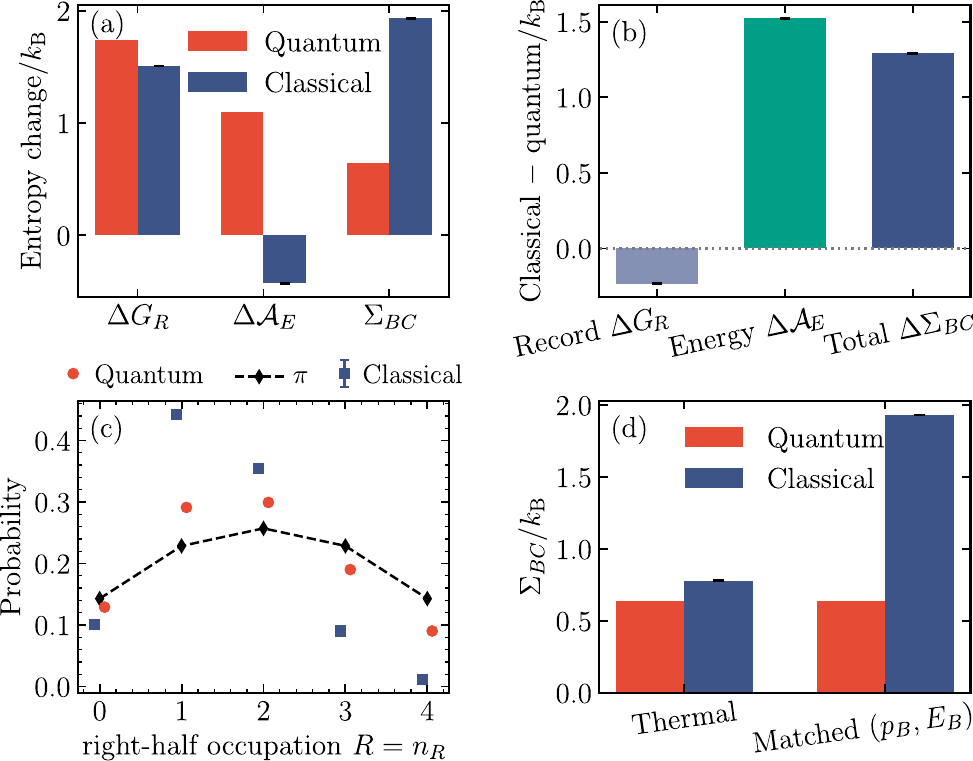}
\caption{
Why the finite-particle quantum process is less irreversible.
(a) Change in the information hidden from the spatial record, the part resolved by adding the mean energy, and the resulting process entropy generation for the strict initial matching.
(b) Classical-minus-quantum process gap split into record and energy-resolved parts. The record part alone favors the classical system, while the energy-resolved part produces the quantum advantage.
(c) Endpoint record distributions and the common multiplicity reference. The quantum record is more relaxed even though the combined energy-and-record description leaves a smaller residual information gap.
(d) Process entropy generation for the natural thermal preparations and for the stricter preparation with matched initial record and mean energy. Classical error bars are standard errors over independent scrambled quasi-Monte Carlo realizations.\justifying}
\label{fig:mechanism}
\end{figure}

Under this stricter preparation,
\(\Sigma_{BC}^{\rm cl,matched}/k_{\rm B}
=1.932489\pm0.002235\), whereas the quantum value is unchanged.
Figure~\figpanel{fig:mechanism}{a} shows the corresponding changes in
\(G_R\), \(\Aenergy\), and \(\Sigma_{BC}\).
The negative classical \(\Delta\Aenergy\) does not mean that
\(\Aenergy\) itself is negative. It means that the mean energy resolves
less of the information hidden from the spatial record at \(C^-\) than
it did initially at \(B\). The resulting gap is
\begin{equation}
\begin{aligned}
\frac{
\Sigma_{BC}^{\rm cl,matched}
-
\Sigma_{BC}^{\rm Q}
}{k_{\rm B}}
&=
\underbrace{-0.232490\pm0.000346}_{\text{record}}
\\
&+
\underbrace{1.523369\pm0.002288}_{\text{energy-resolved}}
\\
&=
1.290879\pm0.002235.
\end{aligned}
\end{equation}
Again, the record-only contribution favors the classical system, while
the energy-resolved contribution produces the smaller quantum entropy
generation [Fig.~\figpanel{fig:mechanism}{b}].

The endpoint record distributions make the same point independently.
For the strict comparison,
\(D[p_{C^-}^{\rm Q}\Vert\pi]=0.026448\), whereas
\(D[p_{C^-}^{\rm cl}\Vert\pi]=0.258922\).
The quantum record is therefore much closer to the common multiplicity
reference [Fig.~\figpanel{fig:mechanism}{c}]: viewed only through the
right-half occupation, the quantum system looks more relaxed.
Nevertheless, once the mean energy is retained as well, its residual
information gap is smaller. The reduction of irreversibility is therefore
not a consequence of weaker macroscopic relaxation.

The physical reason that energy can distinguish states missed by the
spatial record is straightforward. The record \(R\) keeps only the
particle population in the right half, whereas the Hamiltonian also
depends on hopping, the distribution of occupations among individual
sites, relative phases, and interactions within the coarse sectors.
Microscopic states with the same value of the spatial record can
therefore have different mean energies. For the finite-particle dynamics
studied here, imposing that additional energy constraint narrows the
record-compatible quantum set more strongly than it narrows the matched
classical one.

We do not attribute this difference to a single microscopic observable.
The reduction survives when the interaction is removed, and a
time-resolved calculation shows that the quantum--classical difference
in the residual information gap develops during the isolated work stroke
rather than appearing only in the endpoint reconstruction (these checks
are given in Notes~\ref{supp:note-interaction} and
\ref{supp:note-time-resolved} of the SM \cite{SuppMat}).

\textit{Completing the cycle without tomography}.--
We now close the process into a physical cycle using only the retained \((E,R)\) data. The key point is to keep a state quantity, \(\Ires\), separate from process entropy generation, \(\Sigma\).
At \(C^-\), the controller knows the mean energy and the record probabilities but not the full microscopic state. These data determine \(\bar\rho_{E,R}(C^-)\) and its multipliers. From them we define
\begin{equation}
H_{C^-,R}^{\rm Q}
=
k_{\rm B}T_c
\left(
\beta_{C^-}^{\star,{\rm Q}}H_0^{\rm Q}
+
\sum_r\alpha_{r,C^-}^{\rm Q}P_r
\right),
\end{equation}
where \(h_{C^-,R}^{\rm cl}
=
\theta_c
\left[
\beta_{C^-}^{\star,{\rm cl}}h_0
+
\alpha_{C^-}^{\rm cl}(R)
\right]\). The additive energy gauge is fixed once for the work accounting. By construction, the energy-and-record representative at \(C^-\) is Gibbs for this controlled Hamiltonian at \(T_c\).

The working medium is still isolated when the controller switches from \(H_0\) to \(H_{C^-,R}\). Thus \(\rho_{C^+}=\rho_{C^-}\), and similarly classically. The energy-and-record representative is also the same density operator at the two endpoints, although the numerical value of the mean energy changes because the Hamiltonian has changed. The quantum switch work is
\begin{equation}\nonumber
W_{\rm sw}^{\rm Q}
=
k_{\rm B}T_c
\left[
\beta_{C^-}^{\star,{\rm Q}}E_{C^-}^{\rm Q}
+
\sum_r\alpha_{r,C^-}^{\rm Q}p_r(C^-)
\right]
-
E_{C^-}^{\rm Q}.
\end{equation}
The same retained constraints also give
\begin{equation}
\Tr\!\left(H_{C^-,R}^{\rm Q}\rho_{C^+}^{\rm Q}\right)
=
\Tr\!\left[H_{C^-,R}^{\rm Q}\bar\rho_{E,R}^{\rm Q}(C^-)\right],
\end{equation}
with the analogous classical equality. Since the switch is isolated and the representative entropy is unchanged, its retained-description entropy generation is zero.

At \(C^+\), the cold reservoir is attached once and remains attached until \(D\). During \(C^+\rightarrow\bar C\), \(H_{C^-,R}\) is fixed and the actual state relaxes to \(\rho_{\bar C}=\bar\rho_{E,R}(C^-)\). The fixed Hamiltonian means that no work is done. The energy identity above makes the initial and final mean energies equal, so the net heat over this relaxation is zero. The two endpoints have the same retained \((E,R)\) data and therefore the same \(S_{E,R}\). Their record probabilities and energy need not stay fixed at every intermediate time.

The exact microscopic entropy does change. Its increase during \(C^+\rightarrow\bar C\) is \(\Ires(C^-)\). For the present preparation, \(\Ires(C^-)=\Sigma_{BC}\). This is the same irreversibility viewed in two ways: the isolated process \(B\rightarrow C^-\) creates the gap between the actual state and the \((E,R)\) representative while preserving fine entropy, and the later cold relaxation physically removes that hidden distinction. The two appearances are not added.

Without disconnecting the cold reservoir, the controller next varies the Hamiltonian quasistatically from \(H_{C^-,R}\) to \(H_D\). One explicit quantum path is
\begin{equation}
H_{\bar C D}^{\rm Q}(s)
=
(1-s)H_{C^-,R}^{\rm Q}
+
sH_D^{\rm Q},
\quad 0\leq s\leq1.
\end{equation}
The actual state remains Gibbs at \(T_c\), so this stage is reversible. Hence \(Q_c^X/T_c=S_{E,R}^X(\bar C)-S_{E,R}^X(D)=S_{E,R}^X(C^-)-S_{E,R}^X(D)\), where \(Q_c>0\) is the heat rejected by the working medium. Because the preceding relaxation has zero net heat, this is also the net heat rejected during the full cold contact. We do not model the hot- and cold-reservoir contacts with a specific microscopic bath or master equation. Instead, the reversible isotherms are evaluated as quasistatic Gibbs paths, while \(C^+\rightarrow\bar C\) is specified by its thermalizing endpoint at fixed \(H_{C^-,R}\). The isolated stroke \(B\rightarrow C^-\), by contrast, is evolved microscopically. Thus the cycle construction tests the thermodynamic closure under the stated equilibration and quasistatic assumptions, without committing to a particular reservoir model.

At \(D\), the cold reservoir is removed. The isolated scaling of \(H_D\) back to \(H_A\) leaves the microscopic state unchanged and closes the cycle reversibly. The retained-description bookkeeping is then
\begin{equation}\nonumber
\begin{aligned}
\Sigma_{\rm cyc}^{X,(E,R)}
&=
\Sigma_{A\to B}^{X,(E,R)}
+
\Sigma_{B\to C^-}^{X,(E,R)}
+
\Sigma_{C^-\to C^+}^{X,(E,R)}
+
\Sigma_{C^+\to\bar C}^{X,(E,R)}\\
&+
\Sigma_{\bar C\to D}^{X,(E,R)}
+
\Sigma_{D\to A}^{X,(E,R)}
\\
&=
0+\Sigma_{BC}^X+0+0+0+0
=
\Sigma_{BC}^X .
\end{aligned}
\end{equation}
The physical microscopic bookkeeping locates the same entropy generation at the cold relaxation: fine entropy is unchanged during \(B\rightarrow C^-\) and \(C^-\rightarrow C^+\), while \(C^+\rightarrow\bar C\) generates the amount \(\Ires^X(C^-)=\Sigma_{BC}^X\). Thus the retained-description cycle entropy generation and the physical cycle entropy generation are the same number, not two contributions to be added.

The reservoir balance gives the same result, \(-Q_h^X/T_h+Q_c^X/T_c=\Sigma_{BC}^X\). Since the working medium returns to its initial microscopic state and Hamiltonian, \(W_{\rm net}^X=Q_h^X-Q_c^X\), and
\begin{equation}\nonumber
W_{\rm net}^X
=
Q_h^X\left(1-\frac{T_c}{T_h}\right)
-
T_c\Sigma_{BC}^X,
\quad
\eta_X
=
1-\frac{T_c}{T_h}
-
\frac{T_c\Sigma_{BC}^X}{Q_h^X}.
\end{equation}
The factor \(T_c\) comes from the entropy balance of the complete two-reservoir cycle; it is not a temperature assigned to the nonequilibrium state at \(C^-\)~\cite{kondepudi1998modern,degroot1984nonequilibrium,lebellac2004equilibrium,cengel2024thermodynamics,tu2025rethinking}.

The audit verifies the controlled-Hamiltonian energy identity and closes the direct work and entropy balances numerically. For the quantum--classical work comparison we match \(Q_h\), \(T_h\), and \(T_c\) (see Note~\ref{supp:note-cycle} of the SM \cite{SuppMat} for the full state ledger and independent cycle audit). The corresponding efficiency difference is
\begin{equation}
\eta_{\rm Q}-\eta_{\rm cl}
=
\frac{T_c}{Q_h}
\left(
\Sigma_{BC}^{\rm cl}-\Sigma_{BC}^{\rm Q}
\right).
\end{equation}
For the representative four-particle cycle, the finite-particle quantum working medium has the larger efficiency.

\textit{Quantum-to-classical crossover}.--
We next vary the particle number while keeping the classical-field coupling, intensive temperatures, confinement, and expansion time fixed. For each \(Q\), the classical record is rebuilt so that its invariant weights match the quantum multiplicities, and the classical cold scale is adjusted only to match the hot heat input.

For every particle number, the retained \((E,R)\) ledger assigns the only nonzero process term to \(B\rightarrow C^-\). In the physical microscopic cycle, the same amount is generated during the fixed-Hamiltonian relaxation \(C^+\rightarrow\bar C\). The work-only switch, the quasistatic part of the cold contact, the hot isotherm, and the isolated return add no further entropy generation. Therefore \(\Sigma_{\rm cyc}=\Sigma_{BC}\) for every \(Q\).

\begin{figure}[t]
\centering
\includegraphics[width=1\linewidth]{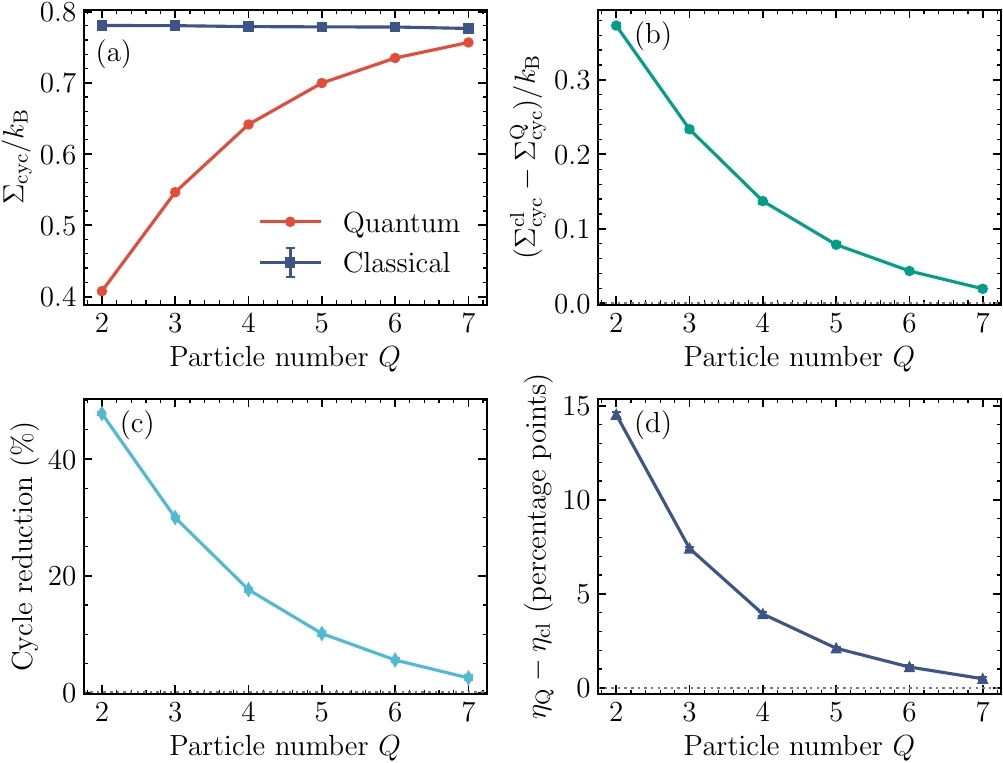}
\caption{
Crossover toward the classical-field regime.
(a) Quantum and classical entropy generation of the complete cycle. The retained \((E,R)\) ledger assigns the nonzero process term to \(B\rightarrow C^-\), while the physical microscopic ledger realizes the same amount during \(C^+\rightarrow\bar C\). The two are not added.
(b) Classical-minus-quantum cycle entropy-generation gap.
(c) Relative reduction of the cycle entropy generation.
(d) Efficiency advantage at matched hot heat input and reservoir temperatures. No asymptotic power law is fitted.\justifying}
\label{fig:crossover}
\end{figure}

Figure~\figpanel{fig:crossover}{a} shows the quantum and classical cycle entropy generation, and Fig.~\figpanel{fig:crossover}{b} shows their difference. The quantum value is lower throughout the computed range, but the gap shrinks as \(Q\) grows. Figure~\figpanel{fig:crossover}{c} shows the same decrease in relative terms, and Fig.~\figpanel{fig:crossover}{d} shows the corresponding efficiency advantage at matched cycle resources. We do not fit a power law. The controlled conclusion is simply that the advantage becomes smaller as the Bose--Hubbard dynamics approaches its classical-field regime (see Note~\ref{supp:note-Q-scaling} of the SM \cite{SuppMat} for he full values and uncertainties).

\textit{Robustness of the complete cycle}.--
We finally vary only the duration of the isolated expansion. The working medium, initial thermal state, spatial record, equilibrium points, and reversible return are kept fixed.

A new expansion time gives a new microscopic endpoint \(C^-\). We therefore recompute its mean energy and record probabilities, rebuild \(\bar\rho_{E,R}(C^-)\), and construct a new \(H_{C^-,R}\) for every duration. The work-only switch is then defined for that endpoint. At \(C^+\), the cold reservoir is attached once: the system first relaxes at fixed \(H_{C^-,R}\) and then, without breaking contact, follows the quasistatic path to the same equilibrium point \(D\). No cold-side representative or control Hamiltonian is reused from another duration.

\begin{figure}[t]
\centering
\includegraphics[width=1\linewidth]{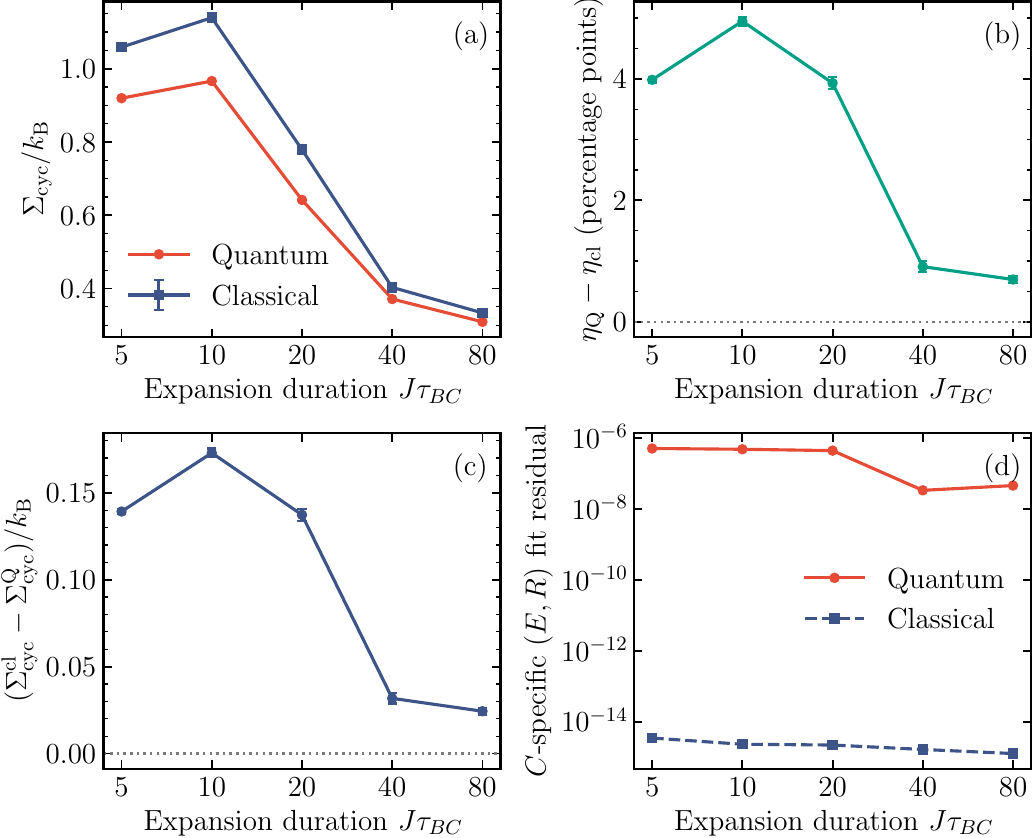}
\caption{
Robustness to the duration of the isolated expansion.
(a) Quantum and classical cycle entropy generation as \(J\tau_{BC}\) is varied.
(b) Efficiency advantage at matched cycle resources.
(c) Classical-minus-quantum cycle entropy-generation gap.
(d) Constraint residual of the energy-and-record reconstruction at each new endpoint \(C^-\). The representative and controlled Hamiltonian are rebuilt for every duration before the cold-side closure is defined.\justifying}
\label{fig:robustness}
\end{figure}

For every duration, the retained \((E,R)\) ledger again gives \(\Sigma_{\rm cyc}=\Sigma_{BC}\), and the physical cold relaxation realizes the same amount. Figure~\figpanel{fig:robustness}{a} shows that the quantum cycle has lower entropy generation at every tested duration. Figures~\figpanel{fig:robustness}{b} and \figpanel{fig:robustness}{c} show the positive efficiency advantage and the cycle gap. Figure~\figpanel{fig:robustness}{d} checks the energy-and-record reconstruction separately at every endpoint. Details are given in Note~\ref{supp:note-protocol-robustness} of the SM \cite{SuppMat}.

We also test the isolated-process mechanism outside the complete-cycle scan. The quantum reduction survives on a larger lattice, remains present without interactions, and persists when the sharp classical partition is replaced by a smooth record derived from the quantum projector (these checks are given in Notes~\ref{supp:note-L6}, \ref{supp:note-interaction}, and \ref{supp:note-soft-record} of the SM \cite{SuppMat}).

\textit{Discussion}.--
The exact microscopic state, the record-only representative, and the energy-and-record representative describe different levels of information. The residual information gap \(\Ires\) is a state quantity: it measures how much microscopic information remains unresolved after both the spatial record and the mean energy are retained. By contrast, \(\Sigma_{BC}\) and \(\Sigma_{\rm cyc}\) are process quantities. During the isolated expansion, the quantum and classical dynamics preserve fine information, but they organize that information differently relative to the same thermodynamic record. In the system studied here, the quantum spatial record relaxes more strongly, while the mean energy still resolves more of the hidden microscopic structure. The endpoint residual gap is therefore smaller, and so is the entropy generation assigned to the isolated work process.

The cycle shows where this same cost appears physically. Using only the retained energy and record, the controller constructs the cold-side Hamiltonian, performs a work-only switch, and then attaches the cold reservoir once. The actual state first relaxes at fixed Hamiltonian, physically realizing the same entropy cost identified from the isolated process, and the same reservoir then remains attached during the reversible cold isotherm. These are two descriptions of one irreversibility, not two separate costs. We do not claim that this ordering is universal: the resolving power of the mean energy depends on the Hamiltonian and the chosen record, and the effect also survives without interactions. Over the particle-number range studied, the quantum advantage decreases as the Bose--Hubbard dynamics approaches its classical-field limit. The result is therefore specific but clear: for this matched finite-particle process, quantum dynamics leaves less microscopic information unusable by the thermodynamic controller, so the same restricted data support a smaller entropy cost, more work, and higher efficiency in the completed cycle.

\section{Acknowledgments}
BA acknowledges support from IRA Program (project no. FENG.02.01-IP.05-0006/23) financed by the FENG program 2021-2027, Priority FENG.02, Measure FENG.02.01., with the support of the FNP.

\bibliography{References}

@misc{SuppMat,
  note = {See Supplemental Material for detailed derivations of the finite-particle quantum--classical correspondence, exact matching of the thermodynamic record, finite-step coarse dynamics, maximum-entropy and Pythagorean constructions, quantum and classical numerical procedures, strict matched-initial-preparation calculation, particle-number crossover, cyclic comparison, protocol-duration and larger-lattice robustness, interaction and soft-record tests, time-resolved accessibility dynamics, numerical error budget, and the complete hierarchy of matching controls.}
}

@book{kondepudi1998modern,
  author    = {Kondepudi, Dilip and Prigogine, Ilya},
  title     = {Modern Thermodynamics: From Heat Engines to Dissipative Structures},
  publisher = {John Wiley \& Sons},
  address   = {New York},
  year      = {1998},
  url       = {https://onlinelibrary.wiley.com/doi/book/10.1002/9781118698723}
}

@book{degroot1984nonequilibrium,
  author    = {de Groot, Sybren R. and Mazur, Peter},
  title     = {Non-Equilibrium Thermodynamics},
  publisher = {Dover Publications},
  address   = {New York},
  year      = {1984},
  url       = {https://api.semanticscholar.org/CorpusID:120634727}
}

@book{lebellac2004equilibrium,
  author    = {Le Bellac, Michel and Mortessagne, Fabrice and Batrouni, G. George},
  title     = {Equilibrium and Non-Equilibrium Statistical Thermodynamics},
  publisher = {Cambridge University Press},
  address   = {Cambridge},
  year      = {2004},
  url       = {https://api.semanticscholar.org/CorpusID:122155388}
}

@book{cengel2024thermodynamics,
  author    = {{\c{C}}engel, Yunus A. and Boles, Michael A. and Kano{\u{g}}lu, Mehmet},
  title     = {Thermodynamics: An Engineering Approach},
  edition   = {10},
  publisher = {McGraw Hill},
  address   = {New York},
  year      = {2024},
  isbn      = {978-1-266-15211-5},
  url       = {https://api.semanticscholar.org/CorpusID:94627517}
}

@article{tu2025rethinking,
  title     = {Rethinking Loss of Available Work and Gouy--Stodola Theorem},
  author    = {Tu, Yaodong and Chen, Gang},
  journal   = {ASME Journal of Heat and Mass Transfer},
  volume    = {147},
  number    = {3},
  pages     = {032901},
  year      = {2025},
  publisher = {American Society of Mechanical Engineers},
  doi       = {10.1115/1.4066860}
}

@article{Trimborn2008,
  author  = {Trimborn, F. and Witthaut, D. and Korsch, H. J.},
  title   = {Exact number-conserving phase-space dynamics of the {$M$}-site Bose--Hubbard model},
  journal = {Phys. Rev. A},
  volume  = {77},
  pages   = {043631},
  year    = {2008},
  doi     = {10.1103/PhysRevA.77.043631}
}

@article{ahmadi2026endogenous,
  title={Thermodynamic Irreversibility from Inaccessible Endogenous Quantum Histories},
  author={Ahmadi, Borhan},
  journal={arXiv:2609.03748},
  year={2026},
  url={https://doi.org/10.48550/arXiv.2609.03748}
}

@book{Huang1987,
  author    = {Huang, Kerson},
  title     = {Statistical Mechanics},
  edition   = {2},
  publisher = {John Wiley \& Sons},
  address   = {New York},
  year      = {1987},
  isbn      = {9780471815181}
}

@book{Reif1965,
  author    = {Reif, Frederick},
  title     = {Fundamentals of Statistical and Thermal Physics},
  publisher = {McGraw-Hill},
  address   = {New York},
  year      = {1965},
  isbn      = {9780070518001}
}

@book{Pathria2011,
  author    = {Pathria, R. K. and Beale, Paul D.},
  title     = {Statistical Mechanics},
  edition   = {3},
  publisher = {Academic Press},
  year      = {2011},
  isbn      = {9780123821881},
  doi       = {10.1016/C2009-0-62310-2}
}

@article{Fisher1989,
  author  = {Fisher, Matthew P. A. and Weichman, Peter B. and Grinstein, G. and Fisher, Daniel S.},
  title   = {Boson Localization and the Superfluid-Insulator Transition},
  journal = {Phys. Rev. B},
  volume  = {40},
  pages   = {546--570},
  year    = {1989},
  doi     = {10.1103/PhysRevB.40.546}
}

@article{Jaksch1998,
  author  = {Jaksch, D. and Bruder, C. and Cirac, J. I. and Gardiner, C. W. and Zoller, P.},
  title   = {Cold Bosonic Atoms in Optical Lattices},
  journal = {Phys. Rev. Lett.},
  volume  = {81},
  pages   = {3108--3111},
  year    = {1998},
  doi     = {10.1103/PhysRevLett.81.3108}
}

@article{Trombettoni2001,
  author  = {Trombettoni, Andrea and Smerzi, Augusto},
  title   = {Discrete Solitons and Breathers with Dilute Bose--Einstein Condensates},
  journal = {Phys. Rev. Lett.},
  volume  = {86},
  pages   = {2353--2356},
  year    = {2001},
  doi     = {10.1103/PhysRevLett.86.2353}
}

@article{Clausius1865,
  author  = {Clausius, Rudolf},
  title   = {Ueber verschiedene f{\"u}r die Anwendung bequeme Formen der Hauptgleichungen der mechanischen W{\"a}rmetheorie},
  journal = {Annalen der Physik},
  volume  = {201},
  number  = {7},
  pages   = {353--400},
  year    = {1865},
  doi     = {10.1002/andp.18652010702},
  url     = {https://doi.org/10.1002/andp.18652010702}
}

@article{Boltzmann1877,
  author  = {Boltzmann, Ludwig},
  title   = {{\"U}ber die Beziehung zwischen dem zweiten Hauptsatze der mechanischen W{\"a}rmetheorie und der Wahrscheinlichkeitsrechnung respektive den S{\"a}tzen {\"u}ber das W{\"a}rmegleichgewicht},
  journal = {Sitzungsberichte der Kaiserlichen Akademie der Wissenschaften in Wien, Mathematisch-Naturwissenschaftliche Classe, Abteilung II},
  volume  = {76},
  pages   = {373--435},
  year    = {1877},
  note    = {English translation with commentary in Entropy 17, 1971--2009 (2015)},
  url     = {https://doi.org/10.3390/e17041971}
}

@book{Gibbs1902,
  author    = {Gibbs, J. Willard},
  title     = {Elementary Principles in Statistical Mechanics: Developed with Especial Reference to the Rational Foundation of Thermodynamics},
  publisher = {Charles Scribner's Sons},
  address   = {New York},
  year      = {1902},
  url       = {https://www.gutenberg.org/ebooks/50992}
}

@article{Jaynes1957,
  author  = {Jaynes, Edwin T.},
  title   = {Information Theory and Statistical Mechanics},
  journal = {Phys. Rev.},
  volume  = {106},
  pages   = {620--630},
  year    = {1957},
  doi     = {10.1103/PhysRev.106.620},
  url     = {https://doi.org/10.1103/PhysRev.106.620}
}

@article{Zwanzig1960,
  author  = {Zwanzig, Robert},
  title   = {Ensemble Method in the Theory of Irreversibility},
  journal = {J. Chem. Phys.},
  volume  = {33},
  number  = {5},
  pages   = {1338--1341},
  year    = {1960},
  doi     = {10.1063/1.1731409},
  url     = {https://doi.org/10.1063/1.1731409}
}

@article{Robertson1966,
  author  = {Robertson, Baldwin},
  title   = {Equations of Motion in Nonequilibrium Statistical Mechanics},
  journal = {Phys. Rev.},
  volume  = {144},
  pages   = {151--161},
  year    = {1966},
  doi     = {10.1103/PhysRev.144.151},
  url     = {https://doi.org/10.1103/PhysRev.144.151}
}

@article{vonNeumann1929,
  author  = {von Neumann, John},
  title   = {Beweis des Ergodensatzes und des {$H$}-Theorems in der neuen Mechanik},
  journal = {Zeitschrift f{\"u}r Physik},
  volume  = {57},
  pages   = {30--70},
  year    = {1929},
  doi     = {10.1007/BF01339852},
  url     = {https://doi.org/10.1007/BF01339852}
}

@article{ReimannEvstigneev2013,
  author  = {Reimann, Peter and Evstigneev, Mykhaylo},
  title   = {Quantum versus Classical Foundation of Statistical Mechanics under Experimentally Realistic Conditions},
  journal = {Phys. Rev. E},
  volume  = {88},
  pages   = {052114},
  year    = {2013},
  doi     = {10.1103/PhysRevE.88.052114},
  url     = {https://doi.org/10.1103/PhysRevE.88.052114}
}

@article{Safranek2019thermo,
  author  = {{\v{S}}afr{\'a}nek, Dominik and Deutsch, J. M. and Aguirre, Anthony},
  title   = {Quantum Coarse-Grained Entropy and Thermodynamics},
  journal = {Phys. Rev. A},
  volume  = {99},
  pages   = {010101},
  year    = {2019},
  doi     = {10.1103/PhysRevA.99.010101},
  url     = {https://doi.org/10.1103/PhysRevA.99.010101}
}

@article{Safranek2020classical,
  author  = {{\v{S}}afr{\'a}nek, Dominik and Aguirre, Anthony and Deutsch, J. M.},
  title   = {Classical Dynamical Coarse-Grained Entropy and Comparison with the Quantum Version},
  journal = {Phys. Rev. E},
  volume  = {102},
  pages   = {032106},
  year    = {2020},
  doi     = {10.1103/PhysRevE.102.032106},
  url     = {https://doi.org/10.1103/PhysRevE.102.032106}
}

@article{Buscemi2023,
  author  = {Buscemi, Francesco and Schindler, Joseph and {\v{S}}afr{\'a}nek, Dominik},
  title   = {Observational Entropy, Coarse-Grained States, and the {Petz} Recovery Map: Information-Theoretic Properties and Bounds},
  journal = {New J. Phys.},
  volume  = {25},
  pages   = {053002},
  year    = {2023},
  doi     = {10.1088/1367-2630/accd11},
  url     = {https://doi.org/10.1088/1367-2630/accd11}
}

@article{Meier2025,
  author  = {Meier, Florian and Rivlin, Tom and Debarba, Tiago and Xuereb, Jake and Huber, Marcus and Lock, Maximilian P. E.},
  title   = {Emergence of a Second Law of Thermodynamics in Isolated Quantum Systems},
  journal = {PRX Quantum},
  volume  = {6},
  pages   = {010309},
  year    = {2025},
  doi     = {10.1103/PRXQuantum.6.010309},
  url     = {https://doi.org/10.1103/PRXQuantum.6.010309}
}

@article{Wilming2016,
  author  = {Wilming, Henrik and Gallego, Rodrigo and Eisert, Jens},
  title   = {Second Law of Thermodynamics under Control Restrictions},
  journal = {Phys. Rev. E},
  volume  = {93},
  pages   = {042126},
  year    = {2016},
  doi     = {10.1103/PhysRevE.93.042126},
  url     = {https://doi.org/10.1103/PhysRevE.93.042126}
}

@article{SafranekRosaBinder2023,
  author  = {{\v{S}}afr{\'a}nek, Dominik and Rosa, Dario and Binder, Felix C.},
  title   = {Work Extraction from Unknown Quantum Sources},
  journal = {Phys. Rev. Lett.},
  volume  = {130},
  pages   = {210401},
  year    = {2023},
  doi     = {10.1103/PhysRevLett.130.210401},
  url     = {https://doi.org/10.1103/PhysRevLett.130.210401}
}

@article{Rubino2026,
  author  = {Rubino, Giulia and Brukner, {\v{C}}aslav and Manzano, Gonzalo},
  title   = {Coarse-Grained Quantum Thermodynamics: Observation-Dependent Quantities, Observation-Independent Laws},
  journal = {Phys. Rev. Research},
  volume  = {8},
  pages   = {023284},
  year    = {2026},
  doi     = {10.1103/gcyn-18tv},
  url     = {https://doi.org/10.1103/gcyn-18tv}
}

@article{PernambucoCeleri2026,
  author  = {Pernambuco, Tiago and C{\'e}leri, Lucas C.},
  title   = {Geometry of Restricted Information: The Case of Quantum Thermodynamics},
  journal = {Phys. Rev. A},
  volume  = {113},
  pages   = {052217},
  year    = {2026},
  doi     = {10.1103/xsqg-xvgc},
  url     = {https://doi.org/10.1103/xsqg-xvgc}
}

@article{3y8f-yrp8,
  title = {Second law of thermodynamics in closed quantum many-body systems},
  author = {Chiba, Yuuya and Yoneta, Yasushi and Hamazaki, Ryusuke and Shimizu, Akira},
  journal = {PRX Quantum},
  pages = {},
  year = {2026},
  month = {Sep},
  publisher = {American Physical Society},
  doi = {10.1103/3y8f-yrp8},
  url = {https://link.aps.org/doi/10.1103/3y8f-yrp8}
}

@article{HokkyoUeda2026,
  author        = {Hokkyo, Akihiro and Ueda, Masahito},
  title         = {Work Extraction Across a Thermodynamic Hierarchy in Quantum Many-Body Systems},
  journal        = {arXiv: 2608.31001 (2026)},
  url           = {https://arxiv.org/abs/2608.31001}
}

@article{WatanabeTakagi2026,
  author  = {Watanabe, Kaito and Takagi, Ryuji},
  title   = {Universal Work Extraction in Quantum Thermodynamics},
  journal = {Nat. Commun.},
  volume  = {17},
  pages   = {1857},
  year    = {2026},
  doi     = {10.1038/s41467-026-69143-3},
  url     = {https://doi.org/10.1038/s41467-026-69143-3}
}

\clearpage
\onecolumngrid


\setcounter{section}{0}
\setcounter{subsection}{0}
\setcounter{equation}{0}
\setcounter{figure}{0}
\setcounter{table}{0}

\renewcommand{\thesection}{\arabic{section}}
\renewcommand{\thesubsection}{\Alph{subsection}}
\renewcommand{\theequation}{S\arabic{equation}}
\renewcommand{\thefigure}{S\arabic{figure}}
\renewcommand{\thetable}{S\arabic{table}}

\makeatletter
\renewcommand{\p@subsection}{\thesection}
\makeatother

\newcommand{\suppnote}[1]{%
    \refstepcounter{section}%
    \setcounter{subsection}{0}%
    \section*{Supplementary Note \arabic{section}. #1}%
    \addcontentsline{toc}{section}{Supplementary Note \arabic{section}. #1}%
}

\newcommand{\suppsubsection}[1]{%
    \refstepcounter{subsection}%
    \subsection*{\Alph{subsection}. #1}%
    \addcontentsline{toc}{subsection}{\Alph{subsection}. #1}%
}

\section*{Supplementary Materials for ``Finite-Particle Quantum Reduction of Thermodynamic Irreversibility''}

This Supplementary Material gives the details behind the main text. We derive the quantum--classical correspondence, match the thermodynamic record, write the common finite-step coarse dynamics, construct the record-only and energy-and-record maximum-entropy representatives, and spell out the cycle closure. The record-only split follows Ref.~\cite{ahmadi2026endogenous}; adding the mean energy gives the finer \((E,R)\) representative used here. During each isolated work process, the finite-\(Q\) quantum state evolves unitarily and the classical field follows measure-preserving Hamiltonian flow, so fine information is not destroyed. Thermal reservoirs enter only in the stated equilibrium parts of the cycle and in the cold-side relaxation used to close it.

\suppnote{Finite-\(Q\) coherent-state correspondence and classical Hamiltonian-field limit}
\label{supp:note-classical-limit}

\suppsubsection{Fixed-particle-number quantum Hilbert space}
\label{supp:fixed-Q-space}

For \(L\) lattice modes containing exactly \(Q\) identical bosons, the many-body Hilbert space is spanned by Fock states \(|n_1,\ldots,n_L\rangle\) with \(n_j\geq0\) and \(\sum_j n_j=Q\). Its dimension is
\begin{equation}
D_Q
=
\binom{Q+L-1}{Q}.
\end{equation}
For the \(L=4\) particle-number scan, \(D_Q=10,20,35,56,84,120\) for \(Q=2,\ldots,7\), respectively.

Within this sector the Bose--Hubbard Hamiltonian is
\begin{equation}
H_Q(\Lambda)
=
-J\sum_{j=1}^{L-1}
\left(
a_j^\dagger a_{j+1}
+
a_{j+1}^\dagger a_j
\right)
+
\frac{U_Q}{2}
\sum_{j=1}^{L}
n_j(n_j-1)
+
\Lambda N_R,
\quad
N_R=\sum_{j>L/2}n_j.
\end{equation}
The total particle-number operator \(N=\sum_jn_j\) commutes with \(H_Q(\Lambda)\) for every \(\Lambda\), including a time-dependent \(\Lambda(t)\). The quantum evolution therefore remains exactly within the fixed-\(Q\) Hilbert sector.

\suppsubsection{Number-conserving coherent-state map}
\label{supp:coherent-state-map}

The natural bridge to the classical Hamiltonian field is provided by the number-conserving \(SU(L)\) coherent state
\begin{equation}
|z;Q\rangle
=
\frac{1}{\sqrt{Q!}}
\left(
\sum_{j=1}^{L}
z_j a_j^\dagger
\right)^Q
|0\rangle,
\quad
\sum_{j=1}^{L}|z_j|^2=1.
\end{equation}
Its one- and two-particle normally ordered moments satisfy \(\langle a_i^\dagger a_j\rangle=Qz_i^\ast z_j\), \(\langle n_j(n_j-1)\rangle=Q(Q-1)|z_j|^4\), and \(\langle N_R\rangle=Qx_R\), where \(x_R=\sum_{j>L/2}|z_j|^2\). It follows that
\begin{equation}
\frac{
\langle z;Q|H_Q(\Lambda)|z;Q\rangle
}{
Q
}
=
-J\sum_{j=1}^{L-1}
\left(
z_j^\ast z_{j+1}
+
z_{j+1}^\ast z_j
\right)
+
\frac{U_Q(Q-1)}{2}
\sum_{j=1}^{L}|z_j|^4
+
\Lambda x_R.
\end{equation}
Choosing \(U_Q=g/(Q-1)\) therefore yields the exact finite-\(Q\) identity
\begin{equation}
\frac{
\langle z;Q|H_Q(\Lambda)|z;Q\rangle
}{
Q
}
=
h_\Lambda(z).
\end{equation}
Thus the finite-\(Q\) quantum Hamiltonian and the classical Hamiltonian field share exactly the same coherent-state energy landscape under the scaling used throughout the paper. This equality does not identify their dynamics: at finite \(Q\), the quantum state evolves in the full many-body Hilbert space, whereas the classical field follows a trajectory on the coherent-state phase space.

\suppsubsection{Classical Hamiltonian dynamics}
\label{supp:dnls-equations}

The normalized classical amplitudes \(z_j\) are defined up to a common phase, so their physical phase space is \(CP^{L-1}\). With the standard complex Hamiltonian structure, \(i\dot z_j=\partial h_{\Lambda(t)}/\partial z_j^\ast\), the equations of motion are
\begin{equation}
i\dot z_j
=
-J
\left(
z_{j-1}
+
z_{j+1}
\right)
+
g|z_j|^2z_j
+
\Lambda(t)\chi_R(j)z_j,
\end{equation}
where missing neighbors are omitted at the open boundaries and \(\chi_R(j)=1\) on the right half and \(0\) on the left. The evolution preserves \(\sum_j|z_j|^2=1\) even for time-dependent \(\Lambda(t)\).

For an ensemble with Liouville density \(f(z,t)\), the corresponding evolution obeys
\begin{equation}
\frac{\partial f}{\partial t}
+
\left\{
f,
h_{\Lambda(t)}
\right\}
=
0.
\end{equation}
Hamiltonian flow is invertible and preserves the invariant phase-space measure \(d\mu\). Therefore the fine Gibbs entropy
\begin{equation}
S_{\rm fine}^{\rm cl}
=
-k_{\rm B}
\int
f(z,t)\ln f(z,t)\,d\mu(z)
\end{equation}
is conserved throughout the isolated work stroke. Classical coarse thermodynamic irreversibility therefore cannot originate from destruction of the underlying Liouville information.

\suppsubsection{Temperature scaling and many-particle limit}
\label{supp:temperature-scaling}

The quantum Hamiltonian is extensive in particle number under the scaling above, whereas \(h_\Lambda\) is an energy per particle. In the particle-number scan we therefore keep the intensive scale \(\theta=k_{\rm B}T/Q\) fixed. Using the exact coherent-state identity,
\begin{equation}
\exp
\left[
-\frac{
\langle z;Q|H_Q(\Lambda)|z;Q\rangle
}{
k_{\rm B}T
}
\right]
=
\exp
\left[
-\frac{
h_\Lambda(z)
}{
\theta
}
\right].
\end{equation}
This relation concerns the matched coherent-state Boltzmann exponent only. The exact finite-\(Q\) quantum Gibbs state remains the many-body operator \(e^{-H_Q/(k_{\rm B}T)}/Z_Q\); it is never replaced by a classical phase-space density.

In the exact number-conserving phase-space formulation of the Bose--Hubbard model, the finite-\(Q\) quantum evolution contains the classical first-order Liouville generator together with nonclassical corrections that vanish in the many-particle limit under the corresponding scaling \cite{Trimborn2008}. The classical Hamiltonian field used here is therefore the controlled many-particle limit of the same microscopic model. This microscopic correspondence does not imply that the derived thermodynamic quantity \(\Sigma_{BC}\), which additionally involves coarse restriction and maximum-entropy projection, must approach its classical value according to a simple \(1/Q\) law.

\suppnote{Exact matching of the quantum and classical thermodynamic records}
\label{supp:note-matched-record}

\suppsubsection{Quantum macrospace multiplicities}
\label{supp:quantum-multiplicity}

Let \(L=2\ell\), with \(\ell\) sites in each half of the chain. For the coarse record \(R=n_R=r\), the right half contains \(r\) bosons and the left half contains \(Q-r\). The numbers of bosonic occupation configurations in the two halves are \(\binom{r+\ell-1}{r}\) and \(\binom{Q-r+\ell-1}{Q-r}\), respectively. Hence the dimension of the quantum macrospace associated with record \(r\) is
\begin{equation}
\Omega_r^{\rm Q}
=
\binom{r+\ell-1}{r}
\binom{Q-r+\ell-1}{Q-r}.
\end{equation}
Summing over all possible right-half occupations and using Vandermonde's identity gives
\begin{equation}
\sum_{r=0}^{Q}
\Omega_r^{\rm Q}
=
\binom{Q+2\ell-1}{Q}
=
D_Q,
\quad
\pi_r^{\rm Q}
=
\frac{
\Omega_r^{\rm Q}
}{
D_Q
}.
\end{equation}
The reference distribution \(\pi^{\rm Q}\) is therefore fixed solely by quantum macrospace multiplicities. It contains no temperature, dynamical, or trajectory-dependent information.

\suppsubsection{Classical invariant distribution of the right-half fraction}
\label{supp:classical-beta}

Under the uniform invariant measure on \(CP^{2\ell-1}\), the site intensities \(y_j=|z_j|^2\) follow the Dirichlet distribution with parameters \((1,\ldots,1)\) on the simplex \(\sum_jy_j=1\). The sum of the \(\ell\) right-half components,
\(x_R=\sum_{j=\ell+1}^{2\ell}y_j\), therefore has the beta distribution
\begin{equation}
x_R
\sim
{\rm Beta}(\ell,\ell),
\quad
w_\ell(x)
=
\frac{
\Gamma(2\ell)
}{
\Gamma(\ell)^2
}
x^{\ell-1}
(1-x)^{\ell-1},
\quad
0\leq x\leq1.
\end{equation}
This is the marginal distribution of \(x_R\) under the invariant measure itself and is independent of the nonequilibrium preparation or subsequent protocol.

\suppsubsection{Matched-\(\pi\) classical partition}
\label{supp:matched-bins}

To represent the same \(Q+1\) coarse outcomes as the quantum record, we partition the classical coordinate \(x_R\) into regions \(A_r=\{z:b_r\leq x_R(z)<b_{r+1}\}\), with \(b_0=0\) and \(b_{Q+1}=1\). Let \(F_\ell(x)=\int_0^xw_\ell(y)\,dy\) be the beta cumulative distribution. The internal boundaries are chosen from
\begin{equation}
F_\ell(b_r)
=
\sum_{m=0}^{r-1}
\pi_m^{\rm Q},
\quad
r=1,\ldots,Q.
\end{equation}
It then follows exactly that
\begin{equation}
\Omega_r^{\rm cl}
\equiv
\mu(A_r)
=
F_\ell(b_{r+1})
-
F_\ell(b_r)
=
\pi_r^{\rm Q},
\quad
\pi_r^{\rm cl}
=
\pi_r^{\rm Q},
\end{equation}
where the invariant classical measure is normalized so that \(\mu(CP^{L-1})=1\). The equality is analytical; finite quadrature introduces only numerical error in estimating these already fixed volumes.

For \(L=4\) and \(Q=4\), the common reference is \(\pi=(5,8,9,8,5)/35\). Since \(x_R\sim{\rm Beta}(2,2)\), the corresponding internal boundaries are
\begin{equation}
b_1=0.2378967658,
\quad
b_2=0.4134203778,
\quad
b_3=0.5865796222,
\quad
b_4=0.7621032342.
\end{equation}
These boundaries depend only on the chosen record and the invariant quantum multiplicities. They are not fitted to either the initial preparation or the nonequilibrium endpoint. This matching removes a finite-\(Q\) counting ambiguity that would otherwise contaminate the quantum--classical comparison; for example, bins centered naively around \(r/Q\) do not reproduce the finite-\(Q\) bosonic macrospace weights exactly.

\suppnote{Quantum and classical finite-step coarse dynamics}
\label{supp:note-coarse-dynamics}

Because the quantum and classical records are matched, their exact microscopic dynamics can be written in the same coarse form. This does not turn either dynamics into a Markov process. It only separates the part generated from the record-only maximum-entropy state from the part carried by microscopic information that the record does not see.

\suppsubsection{Quantum decomposition}
\label{supp:quantum-coarse-dynamics}

For a quantum state with record probabilities \(p_r=\Tr(P_r\rho)\), write
\begin{equation}
\rho
=
\bar\rho_R
+
\chi_R,
\quad
\bar\rho_R
=
\sum_r
\frac{p_r}{\Omega_r^{\rm Q}}P_r,
\quad
\Tr(P_r\chi_R)=0
\ \forall r.
\end{equation}
After an arbitrary unitary step \(U_\tau\), the new record probabilities satisfy
\begin{equation}
p_r'
=
\sum_{r'}
K_{rr'}^{\rm Q}(\tau)p_{r'}
+
r_r^{\rm Q}(\tau),
\quad
K_{rr'}^{\rm Q}(\tau)
=
\frac{
\Tr
\left(
P_rU_\tau P_{r'}U_\tau^\dagger
\right)
}{
\Omega_{r'}^{\rm Q}
},
\quad
r_r^{\rm Q}(\tau)
=
\Tr
\left(
P_rU_\tau\chi_R U_\tau^\dagger
\right).
\end{equation}
Every element of \(K^{\rm Q}\) is nonnegative, and completeness of the projectors gives \(\sum_rK_{rr'}^{\rm Q}=1\). Hence \(K^{\rm Q}\) is a stochastic matrix. Moreover, using the multiplicity reference \(\pi_r=\Omega_r^{\rm Q}/\Omega_{\rm tot}^{\rm Q}\),
\begin{equation}
\begin{aligned}
\left(
K^{\rm Q}\pi
\right)_r
&=
\frac{1}{\Omega_{\rm tot}^{\rm Q}}
\sum_{r'}
\Tr
\left(
P_rU_\tau P_{r'}U_\tau^\dagger
\right)
\\
&=
\frac{
\Tr(P_r)
}{
\Omega_{\rm tot}^{\rm Q}
}
=
\pi_r.
\end{aligned}
\end{equation}
Thus the macro-uniform part of any unitary step preserves the multiplicity reference exactly. The correction \(r^{\rm Q}\) carries the contribution of the unresolved microscopic component \(\chi_R\); in particular, \(\sum_rr_r^{\rm Q}=0\), as required by normalization.

\suppsubsection{Classical decomposition}
\label{supp:classical-coarse-dynamics}

The classical construction is identical in structure. For record probabilities \(p_r=\int_{A_r}f\,d\mu\), write
\begin{equation}
f
=
\bar f_R
+
\chi_R^{\rm cl},
\quad
\bar f_R(z)
=
\sum_r
\frac{p_r}{\Omega_r^{\rm cl}}
\mathbf 1_{A_r}(z),
\quad
\int_{A_r}\chi_R^{\rm cl}\,d\mu=0
\ \forall r.
\end{equation}
Let \(\Phi_\tau\) denote the Hamiltonian flow over the same finite step. Then
\begin{equation}
p_r'
=
\sum_{r'}
K_{rr'}^{\rm cl}(\tau)p_{r'}
+
r_r^{\rm cl}(\tau),
\end{equation}
with
\begin{equation}
K_{rr'}^{\rm cl}(\tau)
=
\frac{1}{\Omega_{r'}^{\rm cl}}
\int_{A_{r'}}
\mathbf 1_{A_r}
\left(
\Phi_\tau z
\right)
d\mu(z),
\quad
r_r^{\rm cl}(\tau)
=
\int
\mathbf 1_{A_r}
\left(
\Phi_\tau z
\right)
\chi_R^{\rm cl}(z)\,d\mu(z).
\end{equation}
Again \(K_{rr'}^{\rm cl}\geq0\) and \(\sum_rK_{rr'}^{\rm cl}=1\). Because the matched classical reference is \(\pi_r=\Omega_r^{\rm cl}\) for the normalized invariant measure, Liouville measure preservation gives
\begin{equation}
\begin{aligned}
\left(
K^{\rm cl}\pi
\right)_r
&=
\sum_{r'}
\int_{A_{r'}}
\mathbf 1_{A_r}
\left(
\Phi_\tau z
\right)
d\mu(z)
\\
&=
\int
\mathbf 1_{A_r}
\left(
\Phi_\tau z
\right)
d\mu(z)
=
\mu(A_r)
=
\pi_r.
\end{aligned}
\end{equation}
The correction also satisfies \(\sum_rr_r^{\rm cl}=0\). Quantum unitarity and classical Liouville invariance therefore generate the same coarse algebraic structure,
\begin{equation}
p'
=
K^X p
+
r^X,
\quad
K^X\pi=\pi,
\quad
X\in\{{\rm Q},{\rm cl}\}.
\end{equation}
The quantum and classical kernels are not equal. Each \(K^X\) and \(r^X\) comes from its own microscopic dynamics and can differ. What is shared is only the stochastic form and the same stationary multiplicity reference \(\pi\).

\suppnote{Maximum entropy, compatibility, and the Pythagorean identity}
\label{supp:note-pythagorean}

The main-text decomposition comes from two nested sets of microscopic states: one set matches the spatial record alone, and the smaller set matches both the record and the mean energy. We prove the quantum and classical relations here.

\suppsubsection{Record-only maximum-entropy representative}
\label{supp:record-maxent}

For fixed record probabilities \(p=\{p_r\}\), define
\begin{equation}
\Cset_R(p)
=
\left\{
\sigma\geq0:
\Tr\sigma=1,\quad
\Tr(P_r\sigma)=p_r
\ \forall r
\right\}.
\end{equation}
Because the \(P_r\) form mutually orthogonal macrospaces, entropy is maximized by removing structure not fixed by their weights and making the state uniform within each active macrospace. The unique maximum-entropy representative is therefore
\begin{equation}
\bar\rho_R
=
\sum_r
\frac{p_r}{\Omega_r^{\rm Q}}P_r,
\quad
S_R
=
-k_{\rm B}
\sum_r
p_r
\ln
\left(
\frac{p_r}{\Omega_r^{\rm Q}}
\right).
\end{equation}
On the active support \(P_{\rm act}=\sum_{r:p_r>0}P_r\),
\begin{equation}
\ln\bar\rho_R
=
\sum_{r:p_r>0}
\ln
\left(
\frac{p_r}{\Omega_r^{\rm Q}}
\right)
P_r.
\end{equation}
Every \(\sigma\in\Cset_R(p)\) has the same expectation value of this operator:
\begin{equation}
\Tr
\left(
\sigma\ln\bar\rho_R
\right)
=
\sum_{r:p_r>0}
p_r
\ln
\left(
\frac{p_r}{\Omega_r^{\rm Q}}
\right).
\end{equation}
In particular, since the exact state \(\rho\) belongs to \(\Cset_R(p)\),
\begin{equation}
k_{\rm B}
D
\left(
\rho
\Vert
\bar\rho_R
\right)
=
S_R
-
S_{\rm vN}
\equiv
G_R.
\end{equation}
The support condition required for this relative entropy is automatic: if \(p_r=0\), positivity of \(\rho\) implies that \(\rho\) has no support in that macrospace.

\suppsubsection{Energy-and-record maximum-entropy representative}
\label{supp:ER-maxent}

Retaining the mean energy \(E=\Tr(H\rho)\) restricts the compatible set further to
\begin{equation}
\Cset_{E,R}(E,p)
=
\left\{
\sigma\in\Cset_R(p):
\Tr(H\sigma)=E
\right\}
\subseteq
\Cset_R(p).
\end{equation}
For an interior optimum on the active support, maximizing \(-\Tr(\sigma\ln\sigma)\) under normalization, energy, and record constraints is equivalent to varying
\begin{equation}
\begin{aligned}
\mathcal L[\sigma]
={}&
-\Tr(\sigma\ln\sigma)
-\lambda_0
\left(
\Tr\sigma-1
\right)
\\
&-
\beta^\star
\left(
\Tr(H\sigma)-E
\right)
-
\sum_r
\alpha_r
\left[
\Tr(P_r\sigma)-p_r
\right].
\end{aligned}
\end{equation}
The stationary condition gives
\begin{equation}
\bar\rho_{E,R}
=
\frac{
\exp
\left[
-\beta^\star H
-
\sum_r\alpha_rP_r
\right]
}{
Z_{E,R}
},
\quad
\ln\bar\rho_{E,R}
=
-\ln Z_{E,R}
-
\beta^\star H
-
\sum_r\alpha_rP_r.
\end{equation}
No commutation between \(H\) and the projectors \(P_r\) is assumed. The exponential is that of the full operator appearing in the exponent. The multipliers are chosen so that \(\bar\rho_{E,R}\) reproduces \(E\) and every \(p_r\); one common shift of all \(\alpha_r\) is redundant because it can be absorbed into \(Z_{E,R}\).

Since \(\rho\) and \(\bar\rho_{E,R}\) have identical normalization, mean energy, and record probabilities,
\begin{equation}
\Tr
\left[
\left(
\rho-\bar\rho_{E,R}
\right)
\ln\bar\rho_{E,R}
\right]
=
0.
\end{equation}
Because they also have the same \(p_r\), the record-only logarithm satisfies
\begin{equation}
\Tr
\left[
\left(
\rho-\bar\rho_{E,R}
\right)
\ln\bar\rho_R
\right]
=
0.
\end{equation}

\suppsubsection{Nested retained--unresolved structure and equality conditions}
\label{supp:nested-retained-unresolved}

The two maximum-entropy representatives correspond to two different levels of retained information and must not be identified with one another. The record-only representative is the linear coarse state
\begin{equation}
\bar\rho_R
=
\cG_R[\rho]
=
\sum_r
\frac{p_r}{\Omega_r^{\rm Q}}P_r,
\end{equation}
and the exact microscopic state can be written
\begin{equation}
\rho
=
\bar\rho_R
+
\chi_R,
\quad
\chi_R
=
\rho-\bar\rho_R.
\end{equation}
Because \(\bar\rho_R\) reproduces every record probability,
\begin{equation}
\Tr(P_r\chi_R)=0
\quad
\forall r.
\end{equation}
This is precisely the retained--unresolved decomposition used in the restricted-record framework of Ref.~\cite{ahmadi2026endogenous}. The unresolved operator \(\chi_R\) contains all microscopic distinctions absent from the spatial record, including nonuniform structure within a macrospace and, when present, inter-macrospace coherences that do not change the current values of \(p_r\).

Retaining the mean energy defines a second exact decomposition,
\begin{equation}
\rho
=
\bar\rho_{E,R}
+
\chi_{E,R},
\quad
\chi_{E,R}
=
\rho-\bar\rho_{E,R}.
\end{equation}
Now the defining constraints imply
\begin{equation}
\Tr(P_r\chi_{E,R})=0
\quad\forall r,
\quad
\Tr(H\chi_{E,R})=0.
\end{equation}
Thus \(\chi_{E,R}\) is invisible to both pieces of retained thermodynamic information.

The refinement supplied by the mean energy is represented by the operator difference
\begin{equation}
\delta_{E|R}
\equiv
\bar\rho_{E,R}
-
\bar\rho_R.
\end{equation}
Subtracting the two exact decompositions gives
\begin{equation}
\chi_R
=
\delta_{E|R}
+
\chi_{E,R}.
\end{equation}
Both terms on the right are invisible to the spatial record,
\begin{equation}
\Tr(P_r\delta_{E|R})=0,
\quad
\Tr(P_r\chi_{E,R})=0,
\end{equation}
but only the second is also energy neutral. Indeed,
\begin{equation}
\Tr(H\delta_{E|R})
=
E-\Tr(H\bar\rho_R),
\quad
\Tr(H\chi_{E,R})=0.
\end{equation}
The operator \(\delta_{E|R}\) therefore supplies exactly the correction needed to move from the state inferred from \(R\) alone to the state inferred from \((E,R)\). The operators \(\delta_{E|R}\), \(\chi_R\), and \(\chi_{E,R}\) are trace-zero Hermitian differences, not density operators by themselves. This operator decomposition is not a convex mixture of physical states.

The corresponding equality conditions follow immediately from strict positivity of relative entropy on a common support:
\begin{equation}
\rho=\bar\rho_R
\iff
D(\rho\Vert\bar\rho_R)=0,
\end{equation}
\begin{equation}
\rho=\bar\rho_{E,R}
\iff
D(\rho\Vert\bar\rho_{E,R})=0,
\end{equation}
and
\begin{equation}
\bar\rho_{E,R}=\bar\rho_R
\iff
D(\bar\rho_{E,R}\Vert\bar\rho_R)=0.
\end{equation}
Hence all three states coincide only in the special case in which the spatial record alone already determines the maximum-entropy state compatible with the exact microscopic state.

Canonical equilibrium illustrates why the three states must remain distinct. Let
\begin{equation}
\rho_{\rm eq}
=
\frac{e^{-\beta H}}{Z}.
\end{equation}
The canonical state maximizes entropy among all states having its mean energy. The set \(\Cset_{E,R}\) is a subset of that fixed-energy set and contains \(\rho_{\rm eq}\), so no state in \(\Cset_{E,R}\) can have a larger entropy. Therefore
\begin{equation}
\bar\rho_{E,R}
=
\rho_{\rm eq}.
\end{equation}
By contrast, \(\bar\rho_R\) generally differs from the canonical state because the spatial record alone need not determine the energy distribution. Thus, generically,
\begin{equation}
\rho_{\rm eq}
=
\bar\rho_{E,R}
\neq
\bar\rho_R.
\end{equation}
This is the structure used at the equilibrium points \(A\), \(B\), and \(D\) of the cyclic construction.

The classical relations are identical in form. Writing
\begin{equation}
f
=
\bar f_R+\chi_R^{\rm cl}
=
\bar f_{E,R}+\chi_{E,R}^{\rm cl}
\end{equation}
gives
\begin{equation}
\int_{A_r}\chi_R^{\rm cl}d\mu=0,
\quad
\int_{A_r}\chi_{E,R}^{\rm cl}d\mu=0,
\quad
\int h\chi_{E,R}^{\rm cl}d\mu=0.
\end{equation}
Defining
\(\delta_{E|R}^{\rm cl}=\bar f_{E,R}-\bar f_R\)
again yields
\(\chi_R^{\rm cl}=\delta_{E|R}^{\rm cl}+\chi_{E,R}^{\rm cl}\).
At a classical canonical equilibrium point,
\(f_{\rm eq}=\bar f_{E,R}\) generically while \(f_{\rm eq}\neq\bar f_R\).

\suppsubsection{Quantum Pythagorean identity}
\label{supp:quantum-pythagorean-proof}

Consider the difference between the two relative-entropy gaps from the exact state to the nested maximum-entropy representatives. Direct subtraction gives
\begin{equation}
\begin{aligned}
&D
\left(
\rho
\Vert
\bar\rho_R
\right)
-
D
\left(
\rho
\Vert
\bar\rho_{E,R}
\right)
\\
&=
\Tr
\left[
\rho
\left(
\ln\bar\rho_{E,R}
-
\ln\bar\rho_R
\right)
\right].
\end{aligned}
\end{equation}
The operator in parentheses is a linear combination of the identity, \(H\), and the projectors \(P_r\). Since \(\rho\) and \(\bar\rho_{E,R}\) agree on the expectation values of all these constrained observables, \(\rho\) can be replaced by \(\bar\rho_{E,R}\) in the final trace. Therefore
\begin{equation}
\begin{aligned}
D
\left(
\rho
\Vert
\bar\rho_R
\right)
&=
D
\left(
\rho
\Vert
\bar\rho_{E,R}
\right)
\\
&\quad+
D
\left(
\bar\rho_{E,R}
\Vert
\bar\rho_R
\right).
\end{aligned}
\end{equation}
The three distances are exactly the three entropy differences used in the main text:
\begin{equation}
\begin{aligned}
k_{\rm B}
D
\left(
\rho
\Vert
\bar\rho_R
\right)
&=
S_R-S_{\rm vN}
=
G_R,
\\
k_{\rm B}
D
\left(
\rho
\Vert
\bar\rho_{E,R}
\right)
&=
S_{E,R}-S_{\rm vN}
=
\Ires,
\\
k_{\rm B}
D
\left(
\bar\rho_{E,R}
\Vert
\bar\rho_R
\right)
&=
S_R-S_{E,R}
=
\Aenergy.
\end{aligned}
\end{equation}
Hence
\begin{equation}
G_R
=
\Aenergy
+
\Ires.
\end{equation}
All three quantities are nonnegative. If some \(p_r\) vanish, the derivation is carried out on the common active support. If the energy-and-record maximum lies on the boundary rather than in the interior, the same identity follows by restriction to its support or, equivalently, by continuity from interior compatible states.

\suppsubsection{Classical Pythagorean identity}
\label{supp:classical-pythagorean-proof}

The classical proof is structurally identical. For a normalized Liouville density \(f\), let \(\bar f_R\) maximize the Gibbs entropy subject to the bin probabilities \(p_r\), and let \(\bar f_{E,R}\) maximize it subject to the same \(p_r\) and mean energy \(e=\int h f\,d\mu\). For an interior solution,
\begin{equation}
\ln\bar f_{E,R}(z)
=
-\ln Z_{E,R}^{\rm cl}
-
\beta^\star h(z)
-
\alpha_{R(z)}.
\end{equation}
Because \(f\) and \(\bar f_{E,R}\) have the same normalization, energy, and bin probabilities, while \(\bar f_R\) depends only on those bin probabilities,
\begin{equation}
\int
\left(
f-\bar f_{E,R}
\right)
\ln\bar f_{E,R}\,d\mu
=
0,
\quad
\int
\left(
f-\bar f_{E,R}
\right)
\ln\bar f_R\,d\mu
=
0.
\end{equation}
Subtracting the corresponding Kullback--Leibler divergences therefore gives
\begin{equation}
D_{\rm KL}
\left(
f
\Vert
\bar f_R
\right)
=
D_{\rm KL}
\left(
f
\Vert
\bar f_{E,R}
\right)
+
D_{\rm KL}
\left(
\bar f_{E,R}
\Vert
\bar f_R
\right).
\end{equation}
With all differential entropies defined relative to the same invariant measure \(d\mu\),
\begin{equation}
G_R^{\rm cl}
=
\Aenergy^{\rm cl}
+
\Ires^{\rm cl},
\end{equation}
where \(G_R^{\rm cl}=S_R^{\rm cl}-S_{\rm fine}^{\rm cl}\), \(\Aenergy^{\rm cl}=S_R^{\rm cl}-S_{E,R}^{\rm cl}\), and \(\Ires^{\rm cl}=S_{E,R}^{\rm cl}-S_{\rm fine}^{\rm cl}\). As in the quantum case, vanishing bin probabilities are handled on the active support, and boundary optima follow by restriction or continuity.

\suppsubsection{Meaning of the three information gaps}
\label{supp:meaning-three-terms}

The three quantities describe one microscopic state at a time. \(G_R\) is the information missing when only the spatial record is kept. \(\Aenergy\) is the part recovered when the mean energy is added. The remainder, \(\Ires\), is the information still missing from the combined \((E,R)\) description. We call \(\Ires\) the residual information gap. It is a state quantity, not an entropy generation. Throughout the paper we reserve \(\Sigma\) for entropy generated along a process or a complete cycle.

The entropy identity has a direct counterpart in the operator split
\begin{equation}
\chi_R
=
\delta_{E|R}
+
\chi_{E,R}.
\end{equation}
Here \(\delta_{E|R}\) is the change in the representative caused by adding the energy constraint, while \(\chi_{E,R}\) is the operator difference that still remains. The related entropy quantities \(\Aenergy\) and \(\Ires\) are relative-entropy gaps, not norms of these operators. The operator pieces therefore need not add in norm; the exact additive statement is the relative-entropy identity derived above.

The words ``resolved'' and ``accessible'' have a limited meaning here. They refer only to how strongly the retained constraints narrow the set of compatible microscopic states. Measuring the mean energy does not reconstruct the microscopic state, and neither \(\Aenergy\) nor \(\Ires\) is a mutual information.

\suppsubsection{A simple example: why the residual information gap vanishes for a canonical state}
\label{supp:canonical-example}

A simple three-level example makes the initial condition used in the main text transparent. Consider a system with energy eigenstates
\(\{|0\rangle,|1\rangle,|2\rangle\}\) and energies
\(0\), \(\varepsilon\), and \(2\varepsilon\), respectively. Choose the inverse temperature such that
\(e^{-\beta\varepsilon}=1/2\). The canonical equilibrium state is then
\begin{equation}
\rho_B
=
\frac{4}{7}|0\rangle\langle0|
+
\frac{2}{7}|1\rangle\langle1|
+
\frac{1}{7}|2\rangle\langle2|.
\end{equation}
Its mean energy is
\begin{equation}
E_B
=
\Tr(H\rho_B)
=
\frac{4}{7}\varepsilon.
\end{equation}

Now suppose that the thermodynamic record is deliberately coarse. It distinguishes only whether the system lies in the two-dimensional subspace spanned by \(|0\rangle\) and \(|1\rangle\), or in the one-dimensional subspace spanned by \(|2\rangle\). The corresponding projectors are
\begin{equation}
P_0
=
|0\rangle\langle0|
+
|1\rangle\langle1|,
\quad
P_1
=
|2\rangle\langle2|.
\end{equation}
For the canonical state, the record probabilities are therefore
\begin{equation}
p_0
=
\Tr(P_0\rho_B)
=
\frac{6}{7},
\quad
p_1
=
\Tr(P_1\rho_B)
=
\frac{1}{7}.
\end{equation}

If only these two probabilities are retained, the record does not tell us how the probability \(6/7\) is divided between \(|0\rangle\) and \(|1\rangle\). The maximum-entropy state compatible with the record alone therefore distributes this probability uniformly inside the unresolved two-dimensional subspace:
\begin{equation}
\bar\rho_R
=
\frac{3}{7}|0\rangle\langle0|
+
\frac{3}{7}|1\rangle\langle1|
+
\frac{1}{7}|2\rangle\langle2|.
\end{equation}
Thus \(\bar\rho_R\neq\rho_B\). The spatial record alone has discarded information about the relative populations of the first two energy levels.

The mean energy restores the missing distinction in this example. Let the probabilities of the three energy levels in a state compatible with both the record and the energy be \(q_0,q_1,q_2\). The record fixes
\begin{equation}
q_0+q_1
=
\frac{6}{7},
\quad
q_2
=
\frac{1}{7},
\end{equation}
while the mean-energy constraint requires
\begin{equation}
\varepsilon q_1
+
2\varepsilon q_2
=
\frac{4}{7}\varepsilon.
\end{equation}
Since \(q_2=1/7\), the energy constraint gives \(q_1=2/7\), and the record normalization then gives \(q_0=4/7\). Hence the maximum-entropy state compatible with both the record and the mean energy is
\begin{equation}
\bar\rho_{E,R}
=
\frac{4}{7}|0\rangle\langle0|
+
\frac{2}{7}|1\rangle\langle1|
+
\frac{1}{7}|2\rangle\langle2|
=
\rho_B.
\end{equation}

The corresponding entropies make the same point numerically:
\begin{equation}
\frac{S_{\rm vN}(\rho_B)}{k_{\rm B}}
=
-\frac{4}{7}\ln\frac{4}{7}
-\frac{2}{7}\ln\frac{2}{7}
-\frac{1}{7}\ln\frac{1}{7}
\simeq
0.95570,
\end{equation}
whereas
\begin{equation}
\frac{S_R}{k_{\rm B}}
=
-2\frac{3}{7}\ln\frac{3}{7}
-\frac{1}{7}\ln\frac{1}{7}
\simeq
1.00424.
\end{equation}
Because \(\bar\rho_{E,R}=\rho_B\),
\begin{equation}
S_{E,R}
=
S_{\rm vN}(\rho_B)
\simeq
0.95570\,k_{\rm B},
\end{equation}
and therefore
\begin{equation}
\Ires(B)
=
S_{E,R}
-
S_{\rm vN}(\rho_B)
=
0.
\end{equation}
At the same time,
\begin{equation}
\mathcal A_E(B)
=
S_R-S_{E,R}
=
G_R(B)
\simeq
0.04854\,k_{\rm B}.
\end{equation}

This example shows the distinction between the two representatives directly. The record-only state \(\bar\rho_R\) differs from the actual canonical state because the coarse record does not resolve the populations inside its first macrospace. Once the mean energy is added, that missing distinction becomes resolvable and the energy-and-record representative returns to the actual canonical state.

The numerical example uses commuting energy and record projectors only to make the arithmetic transparent. The general statement used in the main text does not require \([H,P_r]=0\). A canonical state is already the maximum-entropy state among all states with its mean energy \cite{Pathria2011,Huang1987,Reif1965}. Adding its own record probabilities restricts this set while leaving the canonical state inside it, so no state in the smaller compatibility set can have a larger entropy. Therefore \(\bar\rho_{E,R}(B)=\rho_B\) and \(\Ires(B)=0\) in the general construction as well.

\suppsubsection{No general quantum--classical ordering of energy-resolved accessibility}
\label{supp:no-universal-ordering}

The larger quantum \(\Aenergy\) found in the main text is not a general consequence of using the mean energy as an additional constraint. The mean energy supplies one scalar condition in either description, and how much information it resolves depends on the relation between the Hamiltonian and the chosen record.

The simplest limiting case occurs when the Hamiltonian is already completely determined by the record. If
\begin{equation}
H=\sum_r E_r P_r,
\end{equation}
then the record probabilities themselves fix the mean energy,
\begin{equation}
E=\Tr(H\rho)=\sum_r E_r p_r.
\end{equation}
Adding \(E\) therefore removes no states that were compatible with the record alone. In that case,
\begin{equation}
S_{E,R}=S_R,
\quad
\Aenergy=0.
\end{equation}
The same statement can hold in either a quantum or a classical description.

A small counterexample shows explicitly that the classical resolving power can even be larger than the quantum one. Consider first a four-dimensional quantum system divided into two record sectors,
\begin{equation}
P_0
=
|0,a\rangle\langle0,a|
+
|0,b\rangle\langle0,b|,
\quad
P_1
=
|1,a\rangle\langle1,a|
+
|1,b\rangle\langle1,b|,
\end{equation}
with record probabilities
\begin{equation}
p_0=p_1=\frac12.
\end{equation}
The record-only maximum-entropy state is uniform over the four microscopic states, so
\begin{equation}
S_R^{\rm Q}=k_{\rm B}\ln4.
\end{equation}
Now choose the quantum Hamiltonian
\begin{equation}
H_{\rm Q}
=
\frac{\varepsilon}{2}P_1.
\end{equation}
The mean energy is then already fixed by the record:
\begin{equation}
E_{\rm Q}
=
\frac{\varepsilon}{2}p_1
=
\frac{\varepsilon}{4}.
\end{equation}
Knowing this energy adds no information beyond \(p_0\) and \(p_1\). Hence
\begin{equation}
S_{E,R}^{\rm Q}
=
S_R^{\rm Q}
=
k_{\rm B}\ln4,
\quad
\Aenergy^{\rm Q}=0.
\end{equation}

Now consider a classical toy system with four microscopic configurations carrying the same labels
\((0,a)\), \((0,b)\), \((1,a)\), and \((1,b)\), and the same coarse record \(r=0,1\). Let the record probabilities again be
\begin{equation}
p_0=p_1=\frac12.
\end{equation}
With no further information, the record-only maximum-entropy distribution is again uniform over the four microscopic configurations, giving
\begin{equation}
S_R^{\rm cl}=k_{\rm B}\ln4.
\end{equation}
Assign the classical energies
\begin{equation}
h(r,a)=0,
\quad
h(r,b)=\varepsilon,
\quad
r=0,1.
\end{equation}
and impose the same numerical mean energy as in the quantum example,
\begin{equation}
E_{\rm cl}=\frac{\varepsilon}{4}.
\end{equation}
The energy constraint now carries information that is absent from the record. Since the total probability of the two \(b\) configurations must be \(1/4\), maximum entropy distributes this weight equally between the two record sectors. The energy-and-record maximum-entropy distribution is therefore
\begin{equation}
\bar f_{E,R}^{\rm cl}
=
\left(
\frac38,
\frac18,
\frac38,
\frac18
\right),
\end{equation}
where the entries correspond to
\((0,a)\), \((0,b)\), \((1,a)\), and \((1,b)\). Its entropy is
\begin{equation}
\frac{S_{E,R}^{\rm cl}}{k_{\rm B}}
=
-2
\left[
\frac38\ln\frac38
+
\frac18\ln\frac18
\right]
\simeq
1.255482,
\end{equation}
whereas
\begin{equation}
\frac{S_R^{\rm cl}}{k_{\rm B}}
=
\ln4
\simeq
1.386294.
\end{equation}
Thus
\begin{equation}
\frac{\Aenergy^{\rm cl}}{k_{\rm B}}
=
\frac{S_R^{\rm cl}-S_{E,R}^{\rm cl}}{k_{\rm B}}
\simeq
0.130812
>
\frac{\Aenergy^{\rm Q}}{k_{\rm B}}
=
0.
\end{equation}

The two examples have the same record probabilities, the same record multiplicities, and the same numerical mean energy, yet the energy constraint is more informative in the classical example. Their Hamiltonians are different by design because the purpose of the example is only to disprove any universal statement that mean energy must resolve more information quantum mechanically.

Our Bose--Hubbard comparison is much more restrictive. There the quantum and classical descriptions are derived from the same microscopic Hamiltonian structure, their coherent-state energy landscapes are matched, they undergo the same protocol, and their record multiplicities are matched exactly. In the strict comparison they also begin with the same record probabilities and mean energy. The finding
\begin{equation}
\Delta\Aenergy^{\rm Q}
>
\Delta\Aenergy^{\rm cl}
\end{equation}
is therefore not built into the use of energy as a thermodynamic variable. It is a result of how the matched finite-particle quantum and classical dynamics organize the microscopic information hidden from the record.

\suppnote{Quantum numerical procedure}
\label{supp:note-quantum-numerics}

\suppsubsection{Thermal preparation and continuous-ramp propagation}
\label{supp:quantum-propagation}

All quantum calculations are performed directly in the fixed-\(Q\) Fock basis described in Supplementary Note~\ref{supp:note-classical-limit}. The initial state at \(B\) is the exact canonical state
\begin{equation}
\rho_B^{\rm Q}
=
\frac{
e^{-H_Q(\Lambda_0)/(k_{\rm B}T_h)}
}{
\Tr
e^{-H_Q(\Lambda_0)/(k_{\rm B}T_h)}
},
\end{equation}
obtained by exact diagonalization of \(H_Q(\Lambda_0)\). During the isolated work stroke, the confinement is reduced continuously according to
\begin{equation}
\Lambda(t)
=
\Lambda_0
\left(
1-\frac{t}{\tau_{BC}}
\right),
\quad
0\leq t\leq\tau_{BC}.
\end{equation}
Rather than constructing a product of dense short-time propagators, we factor the initial density matrix as \(\rho_B^{\rm Q}=A_0A_0^\dagger\), for example with \(A_0=(\rho_B^{\rm Q})^{1/2}\), and propagate the amplitude matrix through
\begin{equation}
i\dot A(t)
=
H_Q[\Lambda(t)]A(t).
\end{equation}
The endpoint density matrix is then reconstructed as
\begin{equation}
\rho_{C^-}^{\rm Q}
=
A(\tau_{BC})
A^\dagger(\tau_{BC}).
\end{equation}
The production calculations use the adaptive eighth-order DOP853 integrator with relative tolerance \(2\times10^{-9}\) and absolute tolerance \(2\times10^{-11}\).

As an independent propagation check, the earlier \(Q=4\) calculation based on \(400\) midpoint time slices gave \(\Sigma_{BC}^{\rm Q}/k_{\rm B}=0.6415310\), whereas direct integration of the continuous ramp gives \(0.6416095\). The relative shift is approximately \(1.22\times10^{-2}\%\), far below the quantum--classical difference discussed in the main text.

\suppsubsection{Quantum energy-and-record maximum-entropy solver}
\label{supp:quantum-ER-solver}

At the endpoint \(C^-\), the energy-and-record representative must reproduce the exact quantum mean energy \(E_{C^-}^{\rm Q}=\Tr[H_Q(0)\rho_{C^-}^{\rm Q}]\) and the record probabilities \(p_r(C^-)=\Tr(P_r\rho_{C^-}^{\rm Q})\). We determine its Lagrange multipliers by minimizing the convex dual
\begin{equation}
\Phi
\left(
\beta,\{\alpha_r\}
\right)
=
\ln Z(\beta,\alpha)
+
\beta E_{C^-}^{\rm Q}
+
\sum_r
\alpha_rp_r(C^-),
\end{equation}
with
\begin{equation}
Z(\beta,\alpha)
=
\Tr
\exp
\left[
-\beta H_Q(0)
-
\sum_r\alpha_rP_r
\right].
\end{equation}
The corresponding state is \(\bar\rho_{E,R}=Z^{-1}\exp[-\beta H_Q(0)-\sum_r\alpha_rP_r]\). No commutativity between \(H_Q(0)\) and the \(P_r\) is assumed. The gradient directly measures the mismatch between the target and reconstructed constraints:
\begin{equation}
\frac{\partial\Phi}{\partial\beta}
=
E_{C^-}^{\rm Q}
-
\Tr
\left[
H_Q(0)\bar\rho_{E,R}
\right],
\quad
\frac{\partial\Phi}{\partial\alpha_r}
=
p_r(C^-)
-
\Tr
\left[
P_r\bar\rho_{E,R}
\right].
\end{equation}
Because a common shift of all \(\alpha_r\) is absorbed into the normalization, one record multiplier is fixed as a gauge. The optimization is repeated from multiple initial parameter choices to guard against numerical failures.

Across the complete \(Q=2,\ldots,7\) particle-number scan, the maximum absolute residual among the fitted energy-and-record constraints remains below \(7.4\times10^{-6}\), including at \(Q=7\), where the quantum--classical irreversibility gap is smallest. The numerical drift of the fine von Neumann entropy under the nominally unitary propagation is of order \(10^{-8}\) or smaller and is negligible compared with the statistical uncertainty of the classical calculation.

\suppnote{Classical propagation and quasi-Monte Carlo thermodynamics}
\label{supp:note-classical-numerics}

\suppsubsection{Sampling the invariant phase-space measure}
\label{supp:sobol-sphere}

The classical calculations require integration over the normalized invariant measure \(d\mu\) on \(CP^{L-1}\). We generate low-discrepancy samples by starting from independently scrambled Sobol points in \(2L\) real coordinates, transforming the coordinates to standard Gaussian variables \(\xi_k\), pairing them into \(L\) complex amplitudes, and normalizing:
\begin{equation}
z_j
=
\frac{
\xi_{2j-1}
+
i\xi_{2j}
}{
\sqrt{
\sum_{k=1}^{L}
\left(
\xi_{2k-1}^2
+
\xi_{2k}^2
\right)
}
}.
\end{equation}
The resulting vectors are distributed according to the rotationally invariant measure on the complex unit sphere. All observables used here are invariant under the redundant common phase, so the induced sampling is the required invariant measure on \(CP^{L-1}\).

The primary calculations use eight independently scrambled quasi-Monte Carlo realizations, each containing
\begin{equation}
N_{\rm QMC}
=
2^{16}
=
65536
\end{equation}
phase-space points. Unless stated otherwise, classical uncertainties reported in the main text and Supplementary Material are standard errors across these independent scrambles.

\suppsubsection{Canonical initial density and fine entropy}
\label{supp:classical-canonical}

For the natural thermal comparison, the classical initial density is \(f_B^{\rm cl}(z)\propto e^{-h_{\Lambda_0}(z)/\theta_h}\). At a uniform quadrature point \(z_i\), define \(u_i=e^{-h_{\Lambda_0}(z_i)/\theta_h}\) and the normalized importance weight \(w_i=u_i/\sum_j u_j\). Since the underlying quadrature points represent the normalized invariant measure with equal base weight, the initial fine entropy is estimated as
\begin{equation}
\frac{
S_B^{\rm cl}
}{
k_{\rm B}
}
=
-\sum_i
w_i
\ln
\left(
N_{\rm QMC}w_i
\right).
\end{equation}
This is the discrete quadrature estimator of \(-\int f_B^{\rm cl}\ln f_B^{\rm cl}\,d\mu\) relative to the same normalized invariant measure used throughout the classical calculation.

\suppsubsection{Hamiltonian propagation by symmetric splitting}
\label{supp:classical-strang}

Each classical sample is propagated under the time-dependent DNLS Hamiltonian using a second-order Strang splitting between the on-site and hopping generators. During an on-site half step,
\begin{equation}
z_j
\longrightarrow
\exp
\left[
-i
\left(
g|z_j|^2
+
\Lambda\chi_R(j)
\right)
\frac{\Delta t}{2}
\right]
z_j.
\end{equation}
The hopping part is then evolved exactly according to
\begin{equation}
z
\longrightarrow
e^{-iH_{\rm hop}\Delta t}z,
\end{equation}
followed by the second on-site half step. Here \(H_{\rm hop}\) is the single-particle nearest-neighbor hopping matrix corresponding to the hopping term in \(h_\Lambda\). The production calculations use \(J\Delta t=0.05\). The normalization \(\sum_j|z_j|^2\) is monitored throughout the evolution and its numerical drift remains negligible. Independent time-step refinements are reported below in Supplementary Note~\ref{supp:note-protocol-robustness}.

\suppsubsection{Classical energy-and-record maximum-entropy representative}
\label{supp:classical-ER}

Let \(e_{C^-}^{\rm cl}=\int h_0(z)f_{C^-}(z)\,d\mu(z)\) denote the classical endpoint mean energy per particle, where \(h_0\equiv h_{\Lambda=0}\). Once the endpoint bin probabilities \(p_r(C^-)\) are fixed, the maximum-entropy density compatible with both the record and \(e_{C^-}^{\rm cl}\) has, within each region \(A_r\), the form
\begin{equation}
\bar f_{E,R}(z)
=
p_r
\frac{
e^{-\beta h_0(z)}
}{
Z_r(\beta)
},
\quad
z\in A_r,
\quad
Z_r(\beta)
=
\int_{A_r}
e^{-\beta h_0(z)}
d\mu(z).
\end{equation}
The conditional mean energy in bin \(r\) is
\begin{equation}
e_r(\beta)
=
-\frac{\partial}{\partial\beta}
\ln Z_r(\beta),
\end{equation}
and the single global multiplier \(\beta\) is determined by the endpoint energy constraint
\begin{equation}
\sum_r
p_r(C^-)e_r(\beta)
=
e_{C^-}^{\rm cl}.
\end{equation}
Once this constraint is satisfied, the maximum compatible entropy follows directly:
\begin{equation}
\frac{
S_{E,R}^{\rm cl}(C^-)
}{
k_{\rm B}
}
=
-\sum_r
p_r(C^-)\ln p_r(C^-)
+
\sum_r
p_r(C^-)\ln Z_r(\beta)
+
\beta e_{C^-}^{\rm cl}.
\end{equation}
The natural thermal calculations evaluate the required phase-space integrals using the uniform scrambled-Sobol quadrature described above. The strict matched-initial-preparation calculation is much more concentrated in phase space and therefore uses the deterministic reduction and target-adapted sampling developed next.

\suppnote{Strict matched-initial preparation: high-accuracy calculation}
\label{supp:note-matched-preparation}

The natural quantum and classical canonical states live in different microscopic state spaces, so their initial coarse records need not match exactly. To make sure that the finite-\(Q\) reduction does not come from this initial difference, we also use a stricter classical state that reproduces the quantum record probabilities and mean energy at \(B\).

Because \(h_\Lambda\) is an energy per particle, let \(e^{\rm cl}=\int h_\Lambda f^{\rm cl}\,d\mu\) denote the classical intensive energy and, when useful for direct comparison, define the associated extensive quantity \(E^{\rm cl}\equiv Qe^{\rm cl}\). The strict preparation therefore imposes
\begin{equation}
p_B^{\rm cl}
=
p_B^{\rm Q},
\quad
e_B^{\rm cl}
=
\frac{
E_B^{\rm Q}
}{
Q
},
\end{equation}
or equivalently \(E_B^{\rm cl}=E_B^{\rm Q}\). Among all classical densities satisfying these constraints, we choose the maximum-entropy distribution. This removes differences in the initially accessible record and mean energy while leaving the quantum state, the common reference \(\pi\), and the subsequent work protocol unchanged.

Direct importance sampling from the uniform invariant measure is inefficient for this preparation because the strong confinement \(\Lambda_0=20J\) concentrates its statistical weight into a small region of phase space. We therefore evaluate the equilibrium normalization deterministically and construct a trajectory proposal adapted to the same concentrated distribution.

\suppsubsection{Amplitude--phase representation of \(CP^3\)}
\label{supp:amplitude-phase}

For the representative \(L=4\) system, write \(p_j=|z_j|^2\) and parameterize the amplitudes by
\begin{equation}
p_1=(1-x)u,
\quad
p_2=(1-x)(1-u),
\quad
p_3=xv,
\quad
p_4=x(1-v),
\end{equation}
with \(0\leq x,u,v\leq1\). Here \(x=p_3+p_4=x_R\) is the right-half occupation fraction. After removing the common phase, the three nearest-neighbor phase differences \(\theta_1,\theta_2,\theta_3\) provide the remaining angular coordinates. In these variables the normalized \(CP^3\) measure is
\begin{equation}
d\mu
=
6x(1-x)\,
dx\,du\,dv
\prod_{j=1}^{3}
\frac{d\theta_j}{2\pi},
\end{equation}
and the classical Hamiltonian becomes
\begin{equation}
h_\Lambda
=
\frac{g}{2}
\sum_{j=1}^{4}p_j^2
+
\Lambda x
-
2J
\sum_{j=1}^{3}
\sqrt{p_jp_{j+1}}
\cos\theta_j.
\end{equation}
This representation isolates the large confinement contribution in the single coordinate \(x\) and allows the phase integrations to be performed analytically.

\suppsubsection{Analytical phase reduction and deterministic thermodynamics}
\label{supp:bessel-integration}

For fixed amplitudes, each phase integral is of the form
\begin{equation}
\int_0^{2\pi}
\frac{d\theta}{2\pi}
e^{\kappa\cos\theta}
=
I_0(\kappa),
\end{equation}
where \(I_0\) is the modified Bessel function of the first kind. The six-real-dimensional conditional partition integral in macro-bin \(A_r\) therefore reduces exactly to the three-dimensional integral
\begin{equation}
\begin{aligned}
Z_r(\beta,\Lambda)
={}&
\int_{b_r}^{b_{r+1}}
6x(1-x)\,dx
\int_0^1du
\int_0^1dv
\\
&\times
\exp
\left[
-\beta
\left(
\frac{g}{2}
\sum_{j=1}^{4}p_j^2
+
\Lambda x
\right)
\right]
\\
&\times
\prod_{j=1}^{3}
I_0
\left(
2\beta J\sqrt{p_jp_{j+1}}
\right).
\end{aligned}
\end{equation}
The corresponding conditional phase average required for the hopping contribution is
\begin{equation}
\left\langle
\cos\theta_j
\right\rangle
=
\frac{
I_1(\kappa_j)
}{
I_0(\kappa_j)
},
\quad
\kappa_j
=
2\beta J\sqrt{p_jp_{j+1}},
\end{equation}
where \(I_1\) is the modified Bessel function of the first kind of order one. These expressions permit deterministic evaluation of the bin partition functions, conditional energies, and entropies without sampling the phase variables.

The strict initial inverse temperature is determined by the energy-matching condition
\begin{equation}
\sum_r
p_r^B
e_r
\left(
\beta_B^{\rm cl},\Lambda_0
\right)
=
\frac{
E_B^{\rm Q}
}{
Q
}.
\end{equation}
The resulting value is \(\beta_B^{\rm cl}=4.4798325591\,J^{-1}\), and the associated classical fine entropy is
\begin{equation}
\frac{
S_B^{\rm cl}
}{
k_{\rm B}
}
=
-6.3805447275.
\end{equation}
Its negative value is not pathological. \(S_B^{\rm cl}\) is a differential entropy defined relative to the normalized continuous reference measure \(d\mu\); a rescaling of that measure would shift all corresponding differential entropies by a common additive constant. The entropy differences entering \(\Sigma_{BC}\) are independent of that convention.

\begin{figure*}[t]
\centering
\includegraphics[width=0.6\textwidth]{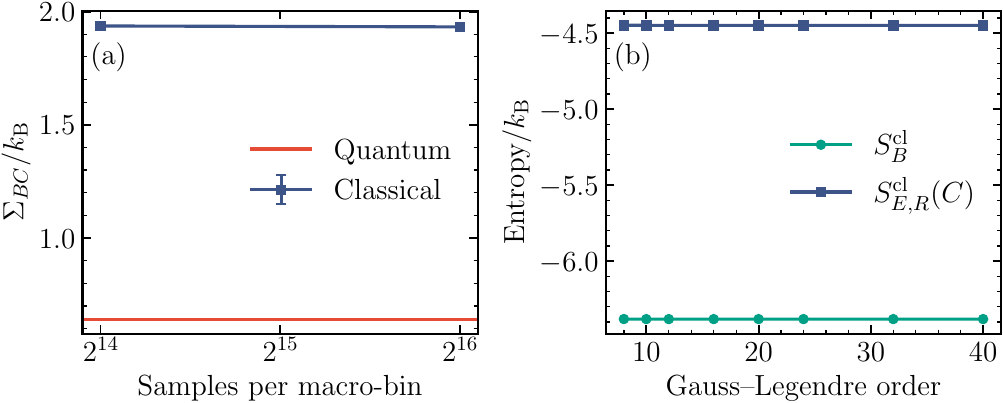}
\caption{
Numerical validation of the strict matched-initial-preparation calculation.
(a) Classical \(\Sigma_{BC}\) obtained with the target-adapted sampler as the number of trajectories per macro-bin is increased from \(2^{14}\) to \(2^{16}\); the horizontal line gives the corresponding quantum value. The classical estimates are statistically consistent as the sampling resolution is increased and remain well separated from the quantum result.
(b) Deterministic Gauss--Legendre convergence of the classical initial fine entropy \(S_B^{\rm cl}\) and of \(S_{E,R}^{\rm cl}(C^-)\) for the fixed representative endpoint used in the quadrature-convergence test. Both quantities are converged to the displayed precision by moderate quadrature order.\justifying}
\label{fig:supp-matched-validation}
\end{figure*}

For the final sampled endpoint constraints, the energy-and-record maximum-entropy calculation gives
\begin{equation}
\frac{
S_{E,R}^{\rm cl}(C^-)
}{
k_{\rm B}
}
=
-4.4480560
\pm
0.0022347,
\end{equation}
and therefore
\begin{equation}
\frac{
\Sigma_{BC}^{\rm cl,matched}
}{
k_{\rm B}
}
=
1.9324887
\pm
0.0022347.
\end{equation}
The quoted uncertainty originates from the sampled endpoint constraints; the deterministic initial entropy is converged to much higher precision. Figure~\figpanel{fig:supp-matched-validation}{b} separately verifies convergence of the deterministic Gauss--Legendre integrations for \(S_B^{\rm cl}\) and for the fixed representative endpoint used in that quadrature-convergence test.

\suppsubsection{Target-adapted trajectory sampling}
\label{supp:target-adapted}

Although the equilibrium normalization and initial entropy can be evaluated deterministically, the endpoint record and energy require propagation of an ensemble of initial conditions. To avoid the poor effective sample size of uniform importance sampling, we choose a proposal that absorbs the dominant confinement dependence,
\begin{equation}
q_x(x)
\propto
x(1-x)
e^{-\beta_B^{\rm cl}\Lambda_0x}.
\end{equation}
At fixed amplitudes, the three phase differences are drawn directly from their exact conditional von Mises distributions,
\begin{equation}
q(\theta_j|p)
=
\frac{
e^{\kappa_j\cos\theta_j}
}{
2\pi I_0(\kappa_j)
},
\quad
\kappa_j
=
2\beta_B^{\rm cl}J
\sqrt{p_jp_{j+1}}.
\end{equation}
The remaining importance weights therefore need to resolve only the residual amplitude-dependent interaction and Bessel factors rather than the dominant confinement and phase concentration.

The production calculation uses \(2^{16}=65536\) trajectories in each of the five macro-bins, giving \(327680\) propagated trajectories per independent realization, with eight independent scrambles. The minimum residual effective-sample fraction is
\begin{equation}
\frac{
N_{\rm eff}
}{
N
}
=
0.64413,
\end{equation}
approximately \(189\) times the corresponding minimum fraction obtained with the original uniform-importance proposal. Figure~\figpanel{fig:supp-matched-validation}{a} shows the stability of the resulting \(\Sigma_{BC}^{\rm cl,matched}\) as the number of trajectories per macro-bin is increased from \(2^{14}\) to \(2^{16}\).

The exact target energy per particle is
\begin{equation}
e_B^{\rm target}
=
\frac{
E_B^{\rm Q}
}{
Q
}
=
0.1428255955J.
\end{equation}
Defining \(\delta e_B=e_B^{\rm sample}-e_B^{\rm target}\), the target-adapted realizations give
\begin{equation}
\delta e_B
=
(-4.02\pm3.06)
\times10^{-4}J.
\end{equation}
After the work stroke, the sampled classical endpoint energy per particle is
\begin{equation}
e_{C^-}^{\rm cl}
=
\frac{
E_{C^-}^{\rm cl}
}{
Q
}
=
-1.12177748
\pm
0.00022407J.
\end{equation}
The corresponding endpoint record is
\begin{equation}
p_{C^-}^{\rm cl}
=
\left(
0.1008555,\,
0.4428753,\,
0.3545321,\,
0.0902859,\,
0.0114512
\right),
\end{equation}
with componentwise standard errors below \(5\times10^{-4}\). Under the same work protocol, the quantum endpoint record is
\begin{equation}
p_{C^-}^{\rm Q}
=
\left(
0.1290675,\,
0.2912160,\,
0.2993206,\,
0.1901327,\,
0.0902632
\right).
\end{equation}
Thus the strict comparison begins with the same accessible record and the same mean energy per particle, yet the two isolated Hamiltonian evolutions arrive at clearly different endpoint records and different energy-and-record irreversibilities. The difference is therefore generated during the dynamics rather than imposed by unequal initial retained data.

\suppnote{Particle-number scaling of the expansion and complete cycle}
\label{supp:note-Q-scaling}

The particle-number scan asks whether the quantum reduction becomes smaller as the Bose--Hubbard dynamics moves toward its classical-field limit. We keep \(g=U_Q(Q-1)=1.3J\), \(\theta_h=1.25J\), \(\theta_c=0.375J\), \(\Lambda_0=20J\), and \(J\tau_{BC}=20\) fixed while varying \(Q=2,\ldots,7\). With this scaling, the classical Hamiltonian and canonical phase-space density do not change with \(Q\). The coarse classical record does change, because it is rebuilt for every particle number so that its invariant weights still match the quantum multiplicities exactly.

For each \(Q\), the cycle is closed by the protocol of Note~\ref{supp:note-cycle} of the SM \cite{SuppMat}. The hot isotherm, the quasistatic part of the cold contact, and the isolated return are reversible. The isolated process \(B\rightarrow C^-\) has process entropy generation \(\Sigma_{BC}^X\), and for the canonical initial state this equals the endpoint state gap \(\Ires^X(C^-)\). The switch \(C^-\rightarrow C^+\) adds no entropy generation. The following cold relaxation physically generates that same amount while the state gap closes. Therefore
\begin{equation}
\Sigma_{\rm cyc}^{X}
=
\Sigma_{BC}^{X},
\quad
X\in\{{\rm Q},{\rm cl}\}.
\end{equation}
No second copy of the same entropy cost is added during cycle closure. For the work comparison, the classical cold scale is adjusted separately at each \(Q\) so that \(Q_h^{\rm cl}=Q_h^{\rm Q}\).

The values used in Fig.~\ref{fig:crossover} are listed in Table~\ref{tab:Q-scaling}. Classical uncertainties are standard errors over independent scrambled quasi-Monte Carlo realizations. Since the later cycle steps add no new sampled entropy cost, the uncertainty of \(\Sigma_{\rm cyc}^{\rm cl}\) is the uncertainty of the classical \(B\rightarrow C^-\) calculation.
\begin{table}[h]
\centering
\caption{Particle-number scaling for the matched-record, \(L=4\) thermal comparison and the completed cycle. For the canonical initial state, the endpoint gap \(\Ires(C^-)\) equals the process entropy generation \(\Sigma_{BC}\). The cold relaxation physically generates this same amount, while the remaining stages add no further entropy generation; hence \(\Sigma_{\rm cyc}=\Sigma_{BC}\). The final column gives the efficiency advantage at matched \(Q_h\), \(T_h\), and \(T_c\).}
\label{tab:Q-scaling}
\begin{tabular}{ccccccc}
\toprule
\(Q\)
&
\(\Sigma_{\rm cyc}^{\rm Q}/k_{\rm B}\)
&
\(\Sigma_{\rm cyc}^{\rm cl}/k_{\rm B}\)
&
\(\Delta\Sigma_{\rm cyc}/k_{\rm B}\)
&
reduction (\%)
&
\(Q_h/J\)
&
\(\Delta\eta\) (pp)
\\
\midrule
2 & 0.407596 & \(0.780501\pm0.003181\) & \(0.372905\pm0.003181\) & \(47.78\pm0.21\) & 1.924993 & \(14.529\pm0.124\) \\
3 & 0.546516 & \(0.780236\pm0.003135\) & \(0.233720\pm0.003135\) & \(29.96\pm0.28\) & 3.546357 & \(7.414\pm0.099\) \\
4 & 0.641610 & \(0.778928\pm0.003298\) & \(0.137318\pm0.003298\) & \(17.63\pm0.35\) & 5.242261 & \(3.929\pm0.094\) \\
5 & 0.699689 & \(0.778537\pm0.003181\) & \(0.078848\pm0.003181\) & \(10.13\pm0.37\) & 7.016720 & \(2.107\pm0.085\) \\
6 & 0.734831 & \(0.778366\pm0.003161\) & \(0.043535\pm0.003161\) & \(5.59\pm0.38\) & 8.829966 & \(1.109\pm0.081\) \\
7 & 0.756688 & \(0.776338\pm0.003225\) & \(0.019649\pm0.003225\) & \(2.53\pm0.40\) & 10.646218 & \(0.484\pm0.080\) \\
\bottomrule
\end{tabular}
\end{table}

The small \(Q\) dependence of the classical entropy generation does not come from a change in the classical Hamiltonian dynamics. It comes from the matched record. As \(Q\) grows, the classical right-half coordinate is divided into more regions so that their invariant weights continue to match the quantum multiplicities. The classical thermodynamic calculation must therefore be repeated for each particle number.

Both the absolute gap and the relative quantum reduction decrease across the scan. At the largest \(Q\) studied, the gap is still larger than the classical sampling uncertainty, but it is much smaller than at low \(Q\). Together with the microscopic correspondence in Note~\ref{supp:note-classical-limit} of the SM \cite{SuppMat}, this supports the finite-particle interpretation. We do not infer an asymptotic power law from this finite range.

The efficiency advantage follows the same trend because the hot heat input is matched and the cycle entropy generation is the same \(\Sigma_{BC}\) identified on the isolated process.

\suppnote{Explicit cycle closure at fixed \texorpdfstring{$Q_h,T_h,T_c$}{Qh, Th, Tc}}
\label{supp:note-cycle}

The isolated process \(B\rightarrow C^-\) is the part of the cycle studied in the main comparison. This Note shows how to close the cycle without reconstructing the full microscopic state. The controller uses only the mean energy, the spatial record probabilities, the two reservoir temperatures, and known Hamiltonian parameters. The reservoirs and the external controller are outside the working-medium boundary. We take \(Q_h>0\) for heat absorbed from the hot reservoir, \(Q_c>0\) for heat rejected to the cold reservoir, \(W_{\rm net}>0\) for net work delivered by the cycle, and \(W_{\rm on}>0\) for work done on the working medium.

The physical sequence is \(A\rightarrow B\rightarrow C^-\rightarrow C^+\rightarrow\bar C\rightarrow D\rightarrow A\). The three labels near \(C\) have different roles. \(C^-\) is the actual nonequilibrium endpoint of the isolated expansion. The switch \(C^-\rightarrow C^+\) changes the Hamiltonian while the system is still isolated, so the microscopic state does not change. The cold reservoir is attached once at \(C^+\) and stays attached until \(D\). During \(C^+\rightarrow\bar C\), the Hamiltonian is fixed and the actual state equilibrates. The same reservoir then remains attached while the Hamiltonian is changed quasistatically during \(\bar C\rightarrow D\).

\suppsubsection{Three-state bookkeeping and equilibrium endpoints}
\label{supp:cycle-three-state-bookkeeping}

At every point we distinguish the exact microscopic state, the record-only maximum-entropy representative, and the energy-and-record maximum-entropy representative. Quantum mechanically these are \(\rho_Y\), \(\bar\rho_R(Y)\), and \(\bar\rho_{E,R}(Y)\); classically they are \(f_Y\), \(\bar f_R(Y)\), and \(\bar f_{E,R}(Y)\). Their relations in the quantum cycle are summarized in Table~\ref{tab:cycle-state-ledger}. The classical equalities are identical after replacing density operators by phase-space densities.

\begin{table}[h]
\centering
\caption{State bookkeeping for the quantum cycle. Equality with the record-only representative is not generic, even at equilibrium. The cold reservoir is attached only once, at \(C^+\), and remains attached through \(D\).}
\label{tab:cycle-state-ledger}

{%
\setlength{\tabcolsep}{8pt}
\renewcommand{\arraystretch}{1.18}

\begin{tabular}{@{}p{0.10\textwidth}p{0.25\textwidth}p{0.27\textwidth}p{0.27\textwidth}@{}}
\toprule
Stage
&
Exact microscopic state
&
Record-only representative
&
Energy-and-record representative
\\
\midrule
\(A\)
&
canonical \(\rho_A\)
&
generally \(\bar\rho_R(A)\neq\rho_A\)
&
\(\bar\rho_{E,R}(A)=\rho_A\)
\\

\(A\rightarrow B\)
&
Gibbs throughout
&
generally different
&
equal to exact state
\\

\(B\)
&
canonical \(\rho_B\)
&
generally \(\bar\rho_R(B)\neq\rho_B\)
&
\(\bar\rho_{E,R}(B)=\rho_B\)
\\

\(B\rightarrow C^-\)
&
unitary \(\rho(t)\)
&
\(\bar\rho_R(t)=\cG_R[\rho(t)]\)
&
generally different from both
\\

\(C^-\)
&
nonequilibrium \(\rho_{C^-}\)
&
\(\bar\rho_R(C^-)\)
&
\(\bar\rho_{E,R}(C^-)\)
\\

\(C^+\)
&
\(\rho_{C^+}=\rho_{C^-}\)
&
\(\bar\rho_R(C^+)=\bar\rho_R(C^-)\)
&
\(\bar\rho_{E,R}(C^+)=\bar\rho_{E,R}(C^-)\)
\\

\(\bar C\)
&
\(\rho_{\bar C}=\bar\rho_{E,R}(C^-)\)
&
\(\bar\rho_R(\bar C)=\bar\rho_R(C^-)\)
&
\(\bar\rho_{E,R}(\bar C)=\rho_{\bar C}\)
\\

\(\bar C\rightarrow D\)
&
Gibbs throughout
&
generally different
&
equal to exact state
\\

\(D\)
&
canonical \(\rho_D\)
&
generally \(\bar\rho_R(D)\neq\rho_D\)
&
\(\bar\rho_{E,R}(D)=\rho_D\)
\\

\(D\rightarrow A\)
&
unchanged equilibrium-family state
&
generally different
&
equal to exact state
\\
\bottomrule
\end{tabular}
}
\end{table}

The table makes the role of the three \(C\)-labels explicit. \(C^-\) and \(C^+\) have the same microscopic state but different Hamiltonians. The point \(\bar C\) is not another inference step: it is the physical point at which the actual state has relaxed to the energy-and-record representative. Because that representative reproduces the record probabilities measured at \(C^-\), the record-only representative is also the same at \(C^-\), \(C^+\), and \(\bar C\), although it is generally different from the actual state at \(\bar C\).

We choose the equilibrium Hamiltonians before following the cycle. The confined hot Hamiltonian is \(H_B^{\rm Q}=H_0^{\rm Q}+\Lambda_0N_R\), with \(H_0^{\rm Q}\equiv H_Q(0)\), and classically \(h_B^{\rm cl}=h_0+\Lambda_0x_R\). The cold equilibrium Hamiltonians are \(H_D^{\rm Q}=\kappa_c^{\rm Q}H_0^{\rm Q}\) and \(h_D^{\rm cl}=\kappa_c^{\rm cl}h_0\). Finally, \(H_A^{\rm Q}=(T_h/T_c)H_D^{\rm Q}\) and \(h_A^{\rm cl}=(T_h/T_c)h_D^{\rm cl}\). Since \(H_A/T_h=H_D/T_c\), the same equilibrium state that is Gibbs at \(D\) is Gibbs at \(A\) after the final isolated energy-scale change.

\suppsubsection{Hot reversible isotherm \texorpdfstring{$A\rightarrow B$}{A to B}}
\label{supp:cycle-AB}

The working medium begins in the canonical state of \(H_A\) at \(T_h\). An explicit quantum path is
\begin{equation}
H_{AB}^{\rm Q}(s)=(1-s)H_A^{\rm Q}+sH_B^{\rm Q},
\quad 0\leq s\leq1,
\end{equation}
with classical counterpart \(h_{AB}^{\rm cl}(s)=(1-s)h_A^{\rm cl}+sh_B^{\rm cl}\). Equivalently, the quantum path has the form \(a_{AB}^{\rm Q}(s)H_0^{\rm Q}+\lambda_{AB}(s)N_R\), where \(a_{AB}^{\rm Q}(s)=(1-s)(T_h/T_c)\kappa_c^{\rm Q}+s\) and \(\lambda_{AB}(s)=s\Lambda_0\).

While the working medium remains quasistatically coupled to the hot reservoir, the actual state is Gibbs for the instantaneous Hamiltonian at \(T_h\). It therefore equals its energy-and-record representative throughout, whereas the record-only representative is generally different. The work done on the quantum working medium is
\begin{equation}
W_{{\rm on},AB}^{\rm Q}
=
\int_0^1ds\,
\Tr\!\left[
\rho_{AB}^{\rm Q}(s)
\frac{\partial H_{AB}^{\rm Q}(s)}{\partial s}
\right],
\end{equation}
with the analogous classical thermodynamic integral. The first law along this explicit equilibrium path gives \(Q_h^X/T_h=S_{E,R}^X(B)-S_{E,R}^X(A)\) and zero entropy generation for \(X\in\{{\rm Q},{\rm cl}\}\).

\suppsubsection{Isolated expansion \texorpdfstring{$B\rightarrow C^-$}{B to C minus}}
\label{supp:cycle-BC}

At \(B\), the exact state is canonical. Therefore \(\rho_B^{\rm Q}=\bar\rho_{E,R}^{\rm Q}(B)\) and \(f_B^{\rm cl}=\bar f_{E,R}^{\rm cl}(B)\), while the record-only representatives are generally different. The residual information gap is zero at this equilibrium point: \(\Ires^X(B)=0\).

The working medium is then isolated and the confinement is lowered according to \(\Lambda(t)=\Lambda_0(1-t/\tau_{BC})\). No heat crosses the working-medium boundary. Quantum mechanically \(\delta W_{\rm on}^{\rm Q}=\langle N_R\rangle d\Lambda\), while classically \(\delta W_{\rm on}^{\rm cl}=Q\langle x_R\rangle d\Lambda\).

The quantum state evolves unitarily and the classical density by measure-preserving Hamiltonian flow, so the fine entropy is constant. At \(C^-\), the quantum Hamiltonian is \(H_0^{\rm Q}\), the exact endpoint is \(\rho_{C^-}^{\rm Q}\), and the retained data are \(E_{C^-}^{\rm Q}=\Tr(H_0^{\rm Q}\rho_{C^-}^{\rm Q})\) and \(p_r(C^-)=\Tr(P_r\rho_{C^-}^{\rm Q})\). The classical endpoint is described by the corresponding mean energy per particle and record probabilities.

We define the record-relative entropy generation of the process \(B\rightarrow C^-\) by
\begin{equation}
\begin{aligned}
\Sigma_{BC}^X
&\equiv
S_{E,R}^X(C^-)-S_{E,R}^X(B)
\\
&=
\Ires^X(C^-)-\Ires^X(B)
=
\Ires^X(C^-).
\end{aligned}
\end{equation}
The first line defines a process quantity. By contrast, \(\Ires^X(C^-)\) is a state quantity: it is the residual information gap of the endpoint. They are equal here because the fine entropy is conserved during the isolated process and \(\Ires^X(B)=0\). For the quantum system, \(\Ires^{\rm Q}(C^-)=k_{\rm B}D[\rho_{C^-}^{\rm Q}\Vert\bar\rho_{E,R}^{\rm Q}(C^-)]\), with the corresponding classical Kullback--Leibler expression. No fine microscopic entropy is produced during \(B\rightarrow C^-\).

\suppsubsection{Work-only record-controlled switch \texorpdfstring{$C^-\rightarrow C^+$}{C minus to C plus}}
\label{supp:cycle-Cswitch}

The exact endpoint \(\rho_{C^-}\) is generally not Gibbs and is never reconstructed. The measured \(E_{C^-}\) and \(p_r(C^-)\) determine the energy-and-record representative. For the quantum system,
\begin{equation}
\bar\rho_{E,R}^{\rm Q}(C^-)
=
\frac{\exp[-\beta_{C^-}^{\star,{\rm Q}}H_0^{\rm Q}-\sum_r\alpha_{r,C^-}^{\rm Q}P_r]}{Z_{C^-}^{\rm Q}}.
\end{equation}
The classical representative has the corresponding form \(\bar f_{E,R}^{\rm cl}\propto\exp[-\beta_{C^-}^{\star,{\rm cl}}h_0-\alpha_{C^-}^{\rm cl}(R)]\). From the same retained quantities we define
\begin{equation}
H_{C^-,R}^{\rm Q}
=
k_{\rm B}T_c
\left[
\beta_{C^-}^{\star,{\rm Q}}H_0^{\rm Q}
+
\sum_r\alpha_{r,C^-}^{\rm Q}P_r
\right],
\quad
h_{C^-,R}^{\rm cl}
=
\theta_c
\left[
\beta_{C^-}^{\star,{\rm cl}}h_0
+
\alpha_{C^-}^{\rm cl}(R)
\right].
\end{equation}
We fix the additive-constant gauge to zero. By construction, \(\bar\rho_{E,R}(C^-)\) is Gibbs for \(H_{C^-,R}^{\rm Q}\) at \(T_c\), and the classical representative is Gibbs for \(h_{C^-,R}^{\rm cl}\). The additional quantum term \(\sum_r\alpha_{r,C^-}P_r\) is a function of \(N_R\), so it requires only the same coarse occupation variable that defines the record.

The switch \(H_0\rightarrow H_{C^-,R}\) is performed while the working medium is isolated and sufficiently rapidly that the microscopic state is unchanged. Thus \(\rho_{C^+}^{\rm Q}=\rho_{C^-}^{\rm Q}\) and \(f_{C^+}^{\rm cl}=f_{C^-}^{\rm cl}\). The quantum switch work is
\begin{equation}
W_{\rm sw}^{\rm Q}
=
k_{\rm B}T_c
\left[
\beta_{C^-}^{\star,{\rm Q}}E_{C^-}^{\rm Q}
+
\sum_r\alpha_{r,C^-}^{\rm Q}p_r(C^-)
\right]
-
E_{C^-}^{\rm Q},
\end{equation}
while classically \(W_{\rm sw}^{\rm cl}=Q\{\theta_c[\beta_{C^-}^{\star,{\rm cl}}e_{C^-}^{\rm cl}+\sum_r\alpha_{r,C^-}^{\rm cl}p_r(C^-)]-e_{C^-}^{\rm cl}\}\). Neither expression contains hidden microscopic variables.

The same constraints give the endpoint identity
\begin{equation}
\Tr\!\left(H_{C^-,R}^{\rm Q}\rho_{C^+}^{\rm Q}\right)
=
\Tr\!\left[H_{C^-,R}^{\rm Q}\bar\rho_{E,R}^{\rm Q}(C^-)\right],
\end{equation}
with the corresponding classical equality. The record probabilities do not change during the switch. The same density operator is also the maximum-entropy representative of the post-switch data, so \(\bar\rho_{E,R}(C^+)=\bar\rho_{E,R}(C^-)\). The numerical mean energy changes because the Hamiltonian changes, but the representative state and its entropy do not. The exact microscopic state and its fine entropy are also unchanged. Hence \(\Ires^X(C^+)=\Ires^X(C^-)\), and the retained-description entropy generation of \(C^-\rightarrow C^+\) is zero.

\suppsubsection{Single cold-reservoir contact \texorpdfstring{$C^+\rightarrow D$}{C plus to D}}
\label{supp:cycle-cold-contact}

At \(C^+\), the cold reservoir at \(T_c\) is attached once and stays attached until \(D\). We split this single contact into two stages because the state is out of equilibrium at \(C^+\) but is Gibbs at \(\bar C\).

\textit{Initial equilibration, \(C^+\rightarrow\bar C\).---}
The controlled Hamiltonian \(H_{C^-,R}\) is held fixed. We assume that the reservoir thermalizes this Hamiltonian. The exact state therefore relaxes to \(\rho_{\bar C}^{\rm Q}=\bar\rho_{E,R}^{\rm Q}(C^-)\) quantum mechanically, and to the corresponding classical Gibbs density classically. No work is done because the Hamiltonian is fixed. The endpoint energy identity makes the initial and final mean energies equal, so the net heat over the relaxation is also zero. Heat can flow back and forth during the relaxation; only the net endpoint transfer is zero.

The retained data have the same values at \(C^+\) and \(\bar C\). The target Gibbs state was built to reproduce the record probabilities at \(C^-\), and the endpoint identity fixes the same mean energy with respect to \(H_{C^-,R}\). The record probabilities and energy need not remain fixed at every intermediate time. At the endpoints, however, \(\bar\rho_{E,R}(\bar C)=\bar\rho_{E,R}(C^+)=\bar\rho_{E,R}(C^-)\) and \(S_{E,R}^X(\bar C)=S_{E,R}^X(C^+)=S_{E,R}^X(C^-)\). The retained-description entropy generation assigned to this segment is therefore zero.

The exact microscopic state does change. At \(C^+\), the residual information gap is \(\Ires^X(C^+)=\Ires^X(C^-)\). At \(\bar C\), the exact state equals its energy-and-record representative, so \(\Ires^X(\bar C)=0\). The fine entropy increase is therefore
\begin{equation}
S_{\rm fine}^X(\bar C)-S_{\rm fine}^X(C^+)
=
\Ires^X(C^-)
=
\Sigma_{BC}^X.
\end{equation}
Because the net heat is zero, this is also the physical entropy generation of the equilibration stage. The equality is important: \(\Ires(C^-)\) is a state gap, \(\Sigma_{BC}\) is the process entropy generation assigned to \(B\rightarrow C^-\), and the cold relaxation physically generates that same amount. They are equal for the present cycle, but they are not the same kind of quantity and they must not be added.

\textit{Quasistatic continuation, \(\bar C\rightarrow D\).---}
Once the state has reached \(\bar C\), it is Gibbs for \(H_{C^-,R}\) at \(T_c\). The cold reservoir remains attached while the controller changes the Hamiltonian quasistatically,
\begin{equation}
H_{\bar C D}^{\rm Q}(s)
=
(1-s)H_{C^-,R}^{\rm Q}+sH_D^{\rm Q},
\quad 0\leq s\leq1.
\end{equation}
The classical path is \(h_{\bar C D}^{\rm cl}(s)=(1-s)h_{C^-,R}^{\rm cl}+sh_D^{\rm cl}\). The actual state stays Gibbs at \(T_c\), so this stage is reversible. Thermodynamic integration gives the work, and \(Q_c^X/T_c=S_{E,R}^X(\bar C)-S_{E,R}^X(D)=S_{E,R}^X(C^-)-S_{E,R}^X(D)\). Here \(Q_c>0\) is heat rejected by the working medium. Since the first stage has zero net heat, this is also the net heat rejected during the full cold contact.

\suppsubsection{Reversible isolated return \texorpdfstring{$D\rightarrow A$}{D to A}}
\label{supp:cycle-DA}

At \(D\), the actual state is canonical at \(T_c\) and equals its energy-and-record representative. We now remove the cold reservoir and thermally isolate the working medium. The Hamiltonian is scaled as \(H_{DA}^{\rm Q}(s)=\gamma(s)H_D^{\rm Q}\), with \(\gamma(0)=1\) and \(\gamma(1)=T_h/T_c\); classically \(h_{DA}^{\rm cl}(s)=\gamma(s)h_D^{\rm cl}\). Since all quantum Hamiltonians along the path are proportional to \(H_D\), the density operator remains unchanged. The classical canonical density is likewise stationary under the scaled Hamiltonian flow. The unchanged state is Gibbs for the instantaneous Hamiltonian at \(T(s)=\gamma(s)T_c\). Hence no heat crosses the boundary, \(S_{E,R}^X(A)=S_{E,R}^X(D)\), and the endpoint is exactly the original equilibrium point \(A\).

\suppsubsection{Two entropy ledgers and the step-by-step cycle balance}
\label{supp:cycle-entropy-ledgers}

There are two useful ways to follow the entropy. The first uses the retained \((E,R)\) description. The second follows the physical fine entropy of the working medium. They place the same irreversibility at different stages, so they must be kept separate.

For the retained description, we write a superscript \((E,R)\) only in this subsection. The step-by-step balance is
\begin{equation}
\begin{aligned}
\Sigma_{\rm cyc}^{X,(E,R)}
={}&
\Sigma_{A\to B}^{X,(E,R)}
+
\Sigma_{B\to C^-}^{X,(E,R)}
+
\Sigma_{C^-\to C^+}^{X,(E,R)}
+
\Sigma_{C^+\to\bar C}^{X,(E,R)}
+
\Sigma_{\bar C\to D}^{X,(E,R)}
+
\Sigma_{D\to A}^{X,(E,R)}
\\
={}&
0
+
\Sigma_{B\to C^-}^{X,(E,R)}
+
0
+
0
+
0
+
0
\\
={}&
\Sigma_{BC}^X,
\quad
X\in\{{\rm Q},{\rm cl}\}.
\end{aligned}
\end{equation}
The zeros come from different facts. \(A\rightarrow B\) and \(\bar C\rightarrow D\) are reversible isotherms. The isolated switch \(C^-\rightarrow C^+\) leaves both the exact state and the entropy of its energy-and-record representative unchanged. The endpoints \(C^+\) and \(\bar C\) have the same retained \((E,R)\) data and zero net heat, so this segment also contributes zero to the retained-description balance. The return \(D\rightarrow A\) is reversible and isolated.

This ledger does not say that fine microscopic entropy is produced during \(B\rightarrow C^-\). The fine entropy is constant there and during the following switch. The physical entropy generation occurs during \(C^+\rightarrow\bar C\). At the start of that relaxation the residual state gap is \(\Ires^X(C^+)=\Ires^X(C^-)\); at \(\bar C\) it has closed to zero. Because the net heat is zero,
\begin{equation}
\Sigma_{C^+\to\bar C}^{X,{\rm phys}}
=
S_{\rm fine}^X(\bar C)-S_{\rm fine}^X(C^+)
=
\Ires^X(C^-)
=
\Sigma_{BC}^X.
\end{equation}
All other stages have zero physical entropy generation. Therefore
\begin{equation}
\Sigma_{\rm cyc}^{X,{\rm phys}}
=
\Sigma_{\rm cyc}^{X,(E,R)}
=
\Sigma_{BC}^X.
\end{equation}
This is one irreversibility written in two ledgers. The isolated process creates the state gap relative to the retained description; the cold relaxation later removes that same hidden distinction. The two amounts are equal here, not additive.

The reservoir entropy balance gives the same cycle result,
\begin{equation}
\Sigma_{\rm cyc}^X
=
-\frac{Q_h^X}{T_h}
+
\frac{Q_c^X}{T_c}
=
\Sigma_{BC}^X.
\end{equation}
We use \(\Sigma_{\rm cyc}^X\) for this common cycle entropy generation from this point on.

The direct work balance is independent of this entropy argument. The total work done on the working medium is
\begin{equation}
W_{{\rm on},{\rm cyc}}^X
=
W_{{\rm on},AB}^X
+
W_{{\rm on},BC}^X
+
W_{\rm sw}^X
+
W_{{\rm on},\bar C D}^X
+
W_{{\rm on},DA}^X,
\end{equation}
because \(C^+\rightarrow\bar C\) has zero work. Since the working medium returns to its initial state and Hamiltonian, \(W_{\rm net}^X=-W_{{\rm on},{\rm cyc}}^X=Q_h^X-Q_c^X\). Eliminating \(Q_c^X\) gives
\begin{equation}
W_{\rm net}^X
=
Q_h^X\left(1-\frac{T_c}{T_h}\right)-T_c\Sigma_{BC}^X,
\quad
\eta_X
=
1-\frac{T_c}{T_h}-\frac{T_c\Sigma_{BC}^X}{Q_h^X}.
\end{equation}
The factor \(T_c\) comes from the entropy balance of the complete two-reservoir cycle. It does not assign a temperature to the nonequilibrium state at \(C^-\) \cite{tu2025rethinking}.

\suppsubsection{Constructive thermodynamic audit of the representative cycle}
\label{supp:cycle-audit}

The audit does not propagate an explicit system--reservoir master equation. The isolated process \(B\rightarrow C^-\) is evolved microscopically, but the hot and cold reversible branches are evaluated from the instantaneous Gibbs states along the specified quasistatic Hamiltonian paths. Likewise, \(C^+\rightarrow\bar C\) is treated through the assumed thermalizing endpoint at fixed \(H_{C^-,R}\), and its zero net heat follows from the endpoint energy identity. The purpose of the audit is therefore to verify the state construction, work and heat accounting, and complete-cycle thermodynamic closure, not the detailed finite-time dynamics of a particular reservoir coupling.

We checked the cycle without using its final entropy or work identities as numerical inputs. For the representative \(L=4\), \(Q=4\), \(g=1.3J\), \(\Lambda_0=20J\), \(k_{\rm B}T_h=5J\), \(k_{\rm B}T_c=1.5J\), and \(J\tau_{BC}=20\) quantum cycle, the exact state at \(B\) was propagated through the isolated ramp to obtain \(\rho_{C^-}\). We then built \(\bar\rho_R(C^-)\) and \(\bar\rho_{E,R}(C^-)\) from that same endpoint. The entropies are \(S_{\rm fine}^{\rm Q}(C^-)/k_{\rm B}=1.603917951140\), \(S_{E,R}^{\rm Q}(C^-)/k_{\rm B}=2.245527508354\), and \(S_R^{\rm Q}(C^-)/k_{\rm B}=3.528900179484\). Thus \(\Ires^{\rm Q}(C^-)/k_{\rm B}=0.641609556920\). Since \(\Ires^{\rm Q}(B)=0\) for the canonical initial state, this is also \(\Sigma_{BC}^{\rm Q}/k_{\rm B}\). The value differs from the production value by only \(2.7\times10^{-7}\) because the audit refits the endpoint representative independently.

The nested decomposition was also checked directly. With \(\chi_R=\rho_{C^-}-\bar\rho_R(C^-)\), \(\delta_{E|R}=\bar\rho_{E,R}(C^-)-\bar\rho_R(C^-)\), and \(\chi_{E,R}=\rho_{C^-}-\bar\rho_{E,R}(C^-)\), the norm of \(\chi_R-\delta_{E|R}-\chi_{E,R}\) is \(3.6\times10^{-18}\). The largest residual of \(\Tr(P_r\chi_R)=0\) is \(2.4\times10^{-17}\), and the dedicated \((E,R)\) fit satisfies the record and energy constraints to \(1.6\times10^{-10}\) or better.

Using the gauge \(\alpha_{4,C^-}=0\), the fitted endpoint multipliers are \(\beta_{C^-}^\star J=0.6261422098\) and \((\alpha_{0,C^-},\alpha_{1,C^-},\alpha_{2,C^-},\alpha_{3,C^-},\alpha_{4,C^-})=(-0.34392177,-0.30243544,-0.05165509,0.18828534,0)\). The representative agrees with the Gibbs state of \(H_{C^-,R}\) at \(T_c\) to \(1.5\times10^{-15}\). The independently evaluated endpoint energies satisfy \(|\Tr(H_{C^-,R}\rho_{C^+})-\Tr[H_{C^-,R}\bar\rho_{E,R}(C^-)]|=1.7\times10^{-10}J\), and the net heat of the fixed-Hamiltonian equilibration is therefore \(-1.7\times10^{-10}J\).

The hot and cold quasistatic stages were evaluated by thermodynamic integration. The resulting work ledger is
\begin{table}[h]
\centering
\caption{Independent work audit for the representative \(L=4\), \(Q=4\) quantum cycle. Negative \(W_{\rm on}\) denotes work delivered by the working medium.}
\label{tab:constructive-cycle-audit}
\begin{tabular}{lc}
\toprule
Quantity & Value \((J)\) \\
\midrule
\(W_{{\rm on},AB}\) & \(45.250322289440\) \\
\(W_{{\rm on},BC}\) & \(-4.373157276174\) \\
\(W_{\rm sw}\) & \(0.062914090983\) \\
\(W_{C^+\bar C}\) & \(0\) \\
\(W_{{\rm on},\bar C D}\) & \(-8.702350833523\) \\
\(W_{{\rm on},DA}\) & \(-34.944896469109\) \\
\midrule
\(Q_h\) & \(5.242260762756\) \\
\(Q_c\) & \(2.535092564207\) \\
\(W_{\rm net}\), direct work sum & \(2.707168198383\) \\
\(W_{\rm net}=Q_h-Q_c\) & \(2.707168198550\) \\
\bottomrule
\end{tabular}
\end{table}
The direct-work closure error is \(-1.7\times10^{-10}J\). Independently, the reservoir entropy balance gives \(-Q_h/T_h+Q_c/T_c=0.641609556920\,k_{\rm B}\), with an entropy-closure error of \(5.8\times10^{-15}k_{\rm B}\). The corresponding quantum efficiency is \(\eta_{\rm Q}=0.516412350\).

A fresh classical Hamiltonian-field ensemble was used for an independent closure audit, not to replace the production ensemble means. For that realized endpoint, \(\beta_{C^-}^\star J=3.3791400672\) and, in the gauge \(\alpha_{4,C^-}=0\), \((\alpha_{0,C^-},\ldots,\alpha_{4,C^-})=(-0.23975696,-0.04811498,0.17045620,0.21214352,0)\). The \(H_{C^-,R}\) energy identity is satisfied to \(5.6\times10^{-16}J\), the net heat of \(C^+\rightarrow\bar C\) is \(-2.2\times10^{-15}J\), and the direct-work and entropy-closure errors are each below \(2.7\times10^{-15}\) in their respective units.

\suppsubsection{Same-\texorpdfstring{$Q_h$}{Qh} matching and representative values}
\label{supp:same-Qh}

For a direct work comparison the quantum and classical cycles use the same hot heat input and reservoir temperatures. We fix \(\kappa_c^{\rm Q}=2.7\) and tune only \(\kappa_c^{\rm cl}\) until the ensemble-mean classical hot input matches the quantum value. This changes the equilibrium points \(D\) and \(A\) of the classical cycle but does not alter the already calculated \(B\rightarrow C^-\) dynamics.

At matched resources, \(W_{\rm net}^{\rm Q}-W_{\rm net}^{\rm cl}=T_c(\Sigma_{BC}^{\rm cl}-\Sigma_{BC}^{\rm Q})\) and \(\eta_{\rm Q}-\eta_{\rm cl}=T_c(\Sigma_{BC}^{\rm cl}-\Sigma_{BC}^{\rm Q})/Q_h\). For \(L=4\), \(Q=4\), \(\kappa_c^{\rm cl}=2.564983\) matches \(Q_h=5.242261J\) at the ensemble-mean level. The cycle entropy generations are \(\Sigma_{\rm cyc}^{\rm Q}/k_{\rm B}=0.641610\) and \(\Sigma_{\rm cyc}^{\rm cl}/k_{\rm B}=0.778928\pm0.003298\), giving \(\eta_{\rm Q}=0.51641\), \(\eta_{\rm cl}=0.47712\), and a quantum efficiency advantage of \(3.929\pm0.094\) percentage points.

\suppsubsection{What the tomography-free closure requires}
\label{supp:cycle-scope}

After \(C^-\), every control parameter is fixed by \(E_{C^-}\), the probabilities \(p_0(C^-),\ldots,p_Q(C^-)\), \(T_c\), \(T_h\), and known Hamiltonian parameters. The protocol does not require the full density matrix, microscopic coherences, phase-space coordinates, or fine occupation patterns. This is the operational role of the thermodynamic reduction here: a small retained record replaces tomography-level information while still fixing the controls needed to close the cycle.

The protocol does require access to the control terms written above. Quantum mechanically, this means an overall scale of \(H_0\) and the record-dependent potential \(F_{C^-}(N_R)=\sum_r\alpha_{r,C^-}P_r\). The classical protocol needs the matching record-dependent potential. This is an explicit assumption about available controls; we do not claim that every platform can implement an arbitrary function of the record without extra engineering.

The cold reservoir is attached only once, at \(C^+\). The relaxation \(C^+\rightarrow\bar C\) and the quasistatic path \(\bar C\rightarrow D\) are two stages of that same uninterrupted contact. The cycle uses the natural canonical preparations used in the particle-number and duration scans. The strict matched-initial classical state of Note~\ref{supp:note-matched-preparation} of the SM \cite{SuppMat} is a separate test of the mechanism and is not the preparation used for the cycle benchmark.

Finally, \(\Ires(C^-)\) and \(\Sigma_{BC}\) must remain conceptually separate. The first is a state property of the nonequilibrium endpoint. The second is the entropy generation assigned to the process \(B\rightarrow C^-\). They are equal in the present cycle because the fine entropy is conserved during the isolated process and the initial canonical state has \(\Ires(B)=0\). The work-only switch adds no entropy generation. The cold relaxation then physically generates the same amount while the residual state gap closes to zero. This is why the retained-description ledger and the physical cycle balance give the same \(\Sigma_{BC}\) without counting it twice.

\suppnote{Protocol-duration robustness and numerical convergence}
\label{supp:note-protocol-robustness}

The duration scan changes only the isolated expansion \(B\rightarrow C^-\). We keep the working medium, thermal preparation, matched spatial record, equilibrium points \(A\), \(B\), and \(D\), and the reversible return \(D\rightarrow A\) fixed. We test \(J\tau_{BC}=5,10,20,40,80\).

Changing \(\tau_{BC}\) changes the actual microscopic endpoint \(\rho_{C^-}\) or \(f_{C^-}\), and therefore changes both thermodynamic representatives \(\bar\rho_R(C^-)\), \(\bar\rho_{E,R}(C^-)\) or their classical analogues. The cold-side construction must therefore be rebuilt from the actual retained data for every duration. For each point in the scan and each microscopic description, we independently recompute the endpoint mean energy and record probabilities, solve the energy-and-record maximum-entropy problem, and construct the corresponding record-controlled Hamiltonian \(H_{C^-,R}(\tau_{BC})\) or \(h_{C^-,R}(\tau_{BC})\).

The physical cold-side continuation is then rebuilt for that endpoint. A work-only switch first carries \(C^-\) to \(C^+\). The cold reservoir is attached once at \(C^+\) and remains attached through \(D\): the system first relaxes at fixed \(H_{C^-,R}(\tau_{BC})\) to \(\bar C\) with zero net heat, and then, without disconnecting the reservoir, the Hamiltonian is varied quasistatically to \(H_D\). No representative, multiplier set, or record-controlled Hamiltonian obtained for one expansion duration is reused for another.

The quantum constraint residual of the endpoint \((E,R)\) reconstructions stays below \(5.2\times10^{-7}\) over the complete scan, while the largest classical residual is below \(3.5\times10^{-15}\). Figure~\figpanel{fig:robustness}{d} displays these endpoint-specific reconstruction checks. They verify that the state \(\bar\rho_{E,R}(C^-)\), or \(\bar f_{E,R}(C^-)\), used to define the cold-side control is recomputed from the actual endpoint constraints for every duration. These residuals test the maximum-entropy reconstruction itself; the zero-net-heat interface and quasistatic cold path follow from the analytical construction in Note~\ref{supp:note-cycle} of the SM \cite{SuppMat}.

The cold equilibrium point \(D\) is fixed throughout the scan. The return \(D\rightarrow A\) is therefore the same reversible Hamiltonian scaling in every row, and the hot isotherm is also unchanged and reversible. For each duration, the canonical initial state has \(\Ires(B)=0\), so the process entropy generation \(\Sigma_{BC}\) equals the endpoint state gap \(\Ires(C^-)\). The initial cold relaxation then physically generates that same amount while the state gap closes. Therefore
\begin{equation}
\Sigma_{\rm cyc}^{X}(\tau_{BC})
=
\Sigma_{BC}^{X}(\tau_{BC}),
\quad
X\in\{{\rm Q},{\rm cl}\}.
\end{equation}
The complete-cycle uncertainty therefore comes directly from the classical expansion calculation rather than from a second sampled return process.

The results are listed in Table~\ref{tab:time-scan}.
\begin{table}[h]
\centering
\caption{Protocol-duration robustness for the matched-record, \(L=4\), \(Q=4\) thermal cycle. The energy-and-record representative and \(H_{C^-,R}\) are rebuilt for every endpoint. For the canonical initial state, \(\Sigma_{BC}=\Ires(C^-)\); the cold relaxation physically generates this same amount, while the remaining stages add no further entropy generation. Thus \(\Sigma_{\rm cyc}=\Sigma_{BC}\) for every row.}
\label{tab:time-scan}

\setlength{\tabcolsep}{9pt}
\renewcommand{\arraystretch}{1.2}

\begin{tabular}{ccccc}
\toprule
\(J\tau_{BC}\)
&
\(\Sigma_{\rm cyc}^{\rm Q}/k_{\rm B}\)
&
\(\Sigma_{\rm cyc}^{\rm cl}/k_{\rm B}\)
&
\(\Delta\Sigma_{\rm cyc}/k_{\rm B}\)
&
\(\Delta\eta\) (pp)
\\
\midrule
5  & 0.919308 & \(1.058532\pm0.001652\) & \(0.139224\pm0.001652\) & \(3.984\pm0.047\) \\
10 & 0.966048 & \(1.138990\pm0.002554\) & \(0.172942\pm0.002554\) & \(4.948\pm0.073\) \\
20 & 0.641610 & \(0.778928\pm0.003298\) & \(0.137318\pm0.003298\) & \(3.929\pm0.094\) \\
40 & 0.371272 & \(0.403029\pm0.002979\) & \(0.031757\pm0.002979\) & \(0.909\pm0.085\) \\
80 & 0.309097 & \(0.333439\pm0.002030\) & \(0.024341\pm0.002030\) & \(0.696\pm0.058\) \\
\bottomrule
\end{tabular}
\end{table}

The quantum cycle entropy remains below the classical value at every tested duration [Fig.~\figpanel{fig:robustness}{a}], and the efficiency advantage remains positive [Fig.~\figpanel{fig:robustness}{b}]. Figure~\figpanel{fig:robustness}{c} shows the corresponding classical-minus-quantum cycle gap. Its dependence on \(\tau_{BC}\) is not monotonic, and no monotonicity claim is made.

\begin{figure*}[t]
\centering
\includegraphics[width=0.7\textwidth]{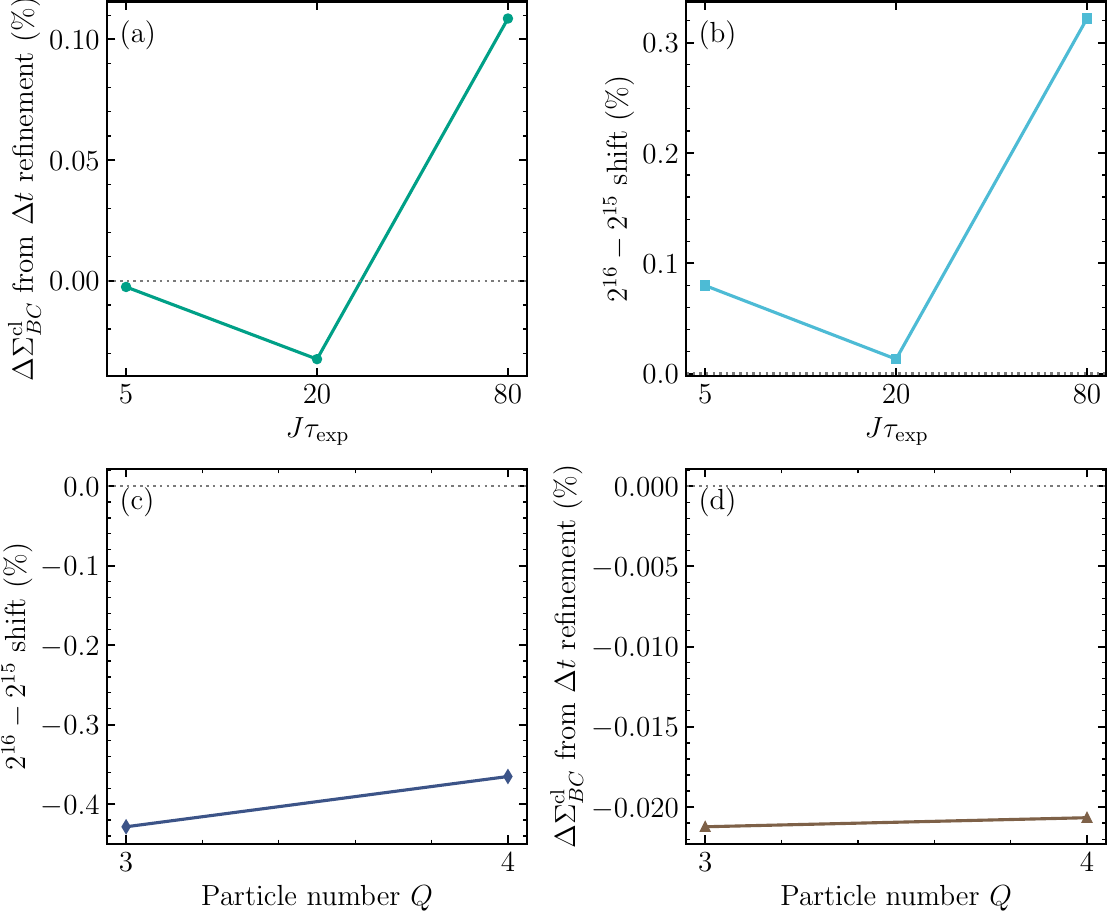}
\caption{
Numerical refinement tests for the isolated expansion calculations.
(a) Relative change of the classical duration-scan results when the split-step time step is reduced.
(b) Relative change of the same results when the Sobol resolution is increased.
(c) Relative change of the larger-lattice results when the Sobol resolution is increased.
(d) Relative change of the larger-lattice results when the classical time step is reduced. The cycle closure requires no additional finite-time simulation of the quasistatic equilibrium strokes; the endpoint-dependent cold-side control is fixed analytically by the retained \((E,R)\) data.\justifying}
\label{fig:supp-refinement}
\end{figure*}

We also refine the two dominant classical numerical resolutions for the isolated expansion. Halving the Strang-splitting step from \(J\Delta t=0.05\) to \(0.025\) changes \(\Sigma_{BC}^{\rm cl}\) by \(-0.0025\%\), \(-0.0324\%\), and \(+0.1086\%\) at \(J\tau_{BC}=5,20,80\), respectively, as shown in Fig.~\figpanel{fig:supp-refinement}{a}. Increasing the scrambled-Sobol resolution from \(2^{15}\) to \(2^{16}\) changes the same points by \(+0.0798\%\), \(+0.0132\%\), and \(+0.3218\%\), respectively [Fig.~\figpanel{fig:supp-refinement}{b}]. These shifts are small compared with the corresponding quantum--classical cycle gaps.

\suppnote{Larger-lattice robustness}
\label{supp:note-L6}

The \(L=6\) calculation tests whether the reduction is specific to the four-site geometry. Each half now contains \(\ell=3\) sites. Under the uniform invariant measure, the classical right-half fraction obeys \(x_R\sim{\rm Beta}(3,3)\), while the quantum multiplicity reference is
\begin{equation}
\pi_r^{\rm Q}
=
\frac{
\binom{r+2}{r}
\binom{Q-r+2}{Q-r}
}{
\binom{Q+5}{Q}
}.
\end{equation}
As in the \(L=4\) construction, the classical boundaries are chosen as beta quantiles of the cumulative quantum weights, giving \(\pi_r^{\rm cl}=\pi_r^{\rm Q}\) exactly in the continuum.

At \(J\tau_{BC}=20\), the production calculations give
\begin{equation}
\begin{aligned}
Q=3:\quad
\frac{\Sigma_{BC}^{\rm Q}}{k_{\rm B}}
&=
0.945160,
\quad
\frac{\Sigma_{BC}^{\rm cl}}{k_{\rm B}}
=
1.259296\pm0.002593,
\\
Q=4:\quad
\frac{\Sigma_{BC}^{\rm Q}}{k_{\rm B}}
&=
1.134840,
\quad
\frac{\Sigma_{BC}^{\rm cl}}{k_{\rm B}}
=
1.255264\pm0.002651.
\end{aligned}
\end{equation}
The corresponding gaps are \(0.314136k_{\rm B}\) and \(0.120424k_{\rm B}\), giving relative quantum reductions of \(24.95\%\) and \(9.59\%\), respectively.

The reversible cycle completion was not evaluated for this \(L=6\) spot check. We therefore use the larger lattice only to test the isolated-stroke ordering \(\Sigma_{BC}^{\rm Q}<\Sigma_{BC}^{\rm cl}\) and do not quote a cyclic efficiency for these points.

The larger phase space also permits independent convergence tests. Increasing the Sobol resolution from \(2^{15}\) to \(2^{16}\) changes the classical \(\Sigma_{BC}\) by \(-0.428\%\) for \(Q=3\) and \(-0.365\%\) for \(Q=4\), as shown in Fig.~\figpanel{fig:supp-refinement}{c}. A further nested \(2^{17}\) calculation using four paired scrambles gives additional changes of only \(-0.050\%\pm0.131\%\) and \(-0.052\%\pm0.166\%\), respectively. Halving the propagation step changes the \(2^{15}\) values by \(-0.0212\%\) and \(-0.0207\%\) [Fig.~\figpanel{fig:supp-refinement}{d}]. These refinements are all small compared with the corresponding gaps \(24.95\%\) and \(9.59\%\) relative to the classical values.

\suppnote{Interaction-strength robustness and microscopic signatures}
\label{supp:note-interaction}

The interaction scan asks whether the reduction needs interaction-generated many-body structure. We keep \(L=4\), \(Q=4\), \(\Lambda_0=20J\), the work protocol, and the strict matched-\((p_B,E_B)\) construction fixed while varying \(g/J=0,0.3,0.7,1.3,2\). At each point \(U_Q=g/(Q-1)\).

The inequality \(\Sigma_{BC}^{\rm Q}<\Sigma_{BC}^{\rm cl}\) persists throughout the complete tested interval. In particular, at \(g=0\) the Bose--Hubbard Hamiltonian is noninteracting, yet
\begin{equation}
\frac{\Sigma_{BC}^{\rm Q}}{k_{\rm B}}
=
0.78235,
\quad
\frac{\Sigma_{BC}^{\rm cl}}{k_{\rm B}}
=
2.24762\pm0.00504.
\end{equation}
The energy-resolved part remains qualitatively different as well: \(\Delta\Aenergy^{\rm Q}/k_{\rm B}=0.97173\), whereas \(\Delta\Aenergy^{\rm cl}/k_{\rm B}=-0.71911\pm0.00408\). The survival of the reduction at \(g=0\) shows that interactions are not necessary for the effect and rules out an explanation that relies exclusively on interaction-generated correlations.

This sweep is a robustness calculation rather than the precision determination of the representative \(g=1.3J\) point. In the lower-resolution scan, the latter gives \(\Sigma_{BC}^{\rm cl}/k_{\rm B}=1.94214\pm0.00255\), whereas the dedicated high-accuracy matched-preparation calculation of Supplementary Note~\ref{supp:note-matched-preparation} gives \(\Sigma_{BC}^{\rm cl,matched}/k_{\rm B}=1.93249\pm0.00223\). The high-accuracy value is used for the quantitative mechanism analysis in the main text; the interaction sweep is used only to establish persistence of the sign and its qualitative dependence on \(g\).

Microscopic observables provide complementary signatures of the distinct finite-\(Q\) dynamics, but they are not used to define the thermodynamic mechanism. At the representative endpoint, the normalized quantum one-body matrix \(\gamma_{ij}^{\rm Q}=\langle a_i^\dagger a_j\rangle/Q\) and its classical-field analogue give approximately \(\Tr[(\gamma^{\rm Q})^2]=0.5538\) and \(\Tr[(\gamma^{\rm cl})^2]=0.7743\), indicating stronger one-body fragmentation on the quantum side. The normalized same-site pair-coincidence values are approximately \(C_2^{\rm Q}=0.2886\) and \(C_2^{\rm cl}=0.3080\). The endpoint hopping contributions per particle are approximately \(e_{\rm hop}^{\rm Q}=-1.1381J\) and \(e_{\rm hop}^{\rm cl}=-1.3226J\), while the interaction contributions are \(e_{\rm int}^{\rm Q}=0.1876J\) and \(e_{\rm int}^{\rm cl}=0.2002J\). These observables characterize microscopic differences between the two dynamics, but none alone establishes causality. The exact thermodynamic statement remains the Pythagorean decomposition \(\Sigma_{BC}=\Delta G_R-\Delta\Aenergy\).

\suppnote{Soft-record check from the quantum projector symbol}
\label{supp:note-soft-record}

The main construction uses disjoint classical regions \(A_r\) whose invariant volumes match the quantum macrospace multiplicities. To check whether the reduction depends on this sharp partition, we also build a smooth classical record from the coherent-state symbol of the quantum number projector.

For a number-conserving coherent state with right-half fraction \(x\), the probability associated with the quantum outcome \(r\) is
\begin{equation}
M_r(x)
=
\langle z;Q|P_r|z;Q\rangle
=
\binom{Q}{r}
x^r
(1-x)^{Q-r}.
\end{equation}
The functions \(M_r(x)\) are nonnegative and satisfy \(\sum_rM_r(x)=1\), but they overlap in phase space rather than defining mutually exclusive regions. For \(L=4\), averaging them over the invariant density \(6x(1-x)\) gives
\begin{equation}
\begin{aligned}
\pi_r^{\rm soft}
&=
\int_0^1
6x(1-x)
M_r(x)\,dx
\\
&=
\frac{
6(r+1)(Q-r+1)
}{
(Q+1)(Q+2)(Q+3)
}
=
\frac{
\Omega_r^{\rm Q}
}{
\Omega_{\rm tot}^{\rm Q}
}
=
\pi_r^{\rm Q},
\end{aligned}
\end{equation}
where \(\Omega_r^{\rm Q}=(r+1)(Q-r+1)\) and \(\Omega_{\rm tot}^{\rm Q}=\binom{Q+3}{3}\). Thus the soft construction reproduces the same quantum multiplicity reference exactly without introducing hard classical bin boundaries.

For a classical density \(f\), the corresponding accessible probabilities are \(p_r=\int M_r(x_R)f(z)\,d\mu(z)\). Because this stochastic record differs structurally from the disjoint partition used in the principal comparison, we use it only as an alternative-record robustness test and not as a replacement for the primary matched-\(\pi\) construction.

Repeating the strict \(Q=4\) classical maximum-entropy analysis with these soft constraints gives
\begin{equation}
\frac{
\Sigma_{BC,\rm soft}^{\rm cl}
}{
k_{\rm B}
}
=
4.0895
\pm
0.0040.
\end{equation}
The finite-\(Q\) quantum reduction therefore survives when the hard classical partition is replaced by the exact coherent-state response function of the quantum projector. The numerical value changes strongly, as expected when the thermodynamic record itself is changed; the significance of this test is the persistence of the ordering, not equality of the hard- and soft-record entropy values.

\suppnote{How the residual information gap develops during the work process}
\label{supp:note-time-resolved}

The final quantum--classical difference does not appear only at \(C^-\). It builds up during the isolated work process. To see this, we carried out a separate time-resolved calculation for the strict \(L=4\), \(Q=4\) matched preparation. This calculation is only a diagnostic check; the endpoint values used in the main text come from the higher-accuracy production calculations.

The residual information gap \(\Ires\) starts to separate early in the ramp. At approximately \(Jt=1\), the diagnostic calculation gives \(\Ires^{\rm Q}/k_{\rm B}\simeq4.9\times10^{-5}\) and \(\Ires^{\rm cl}/k_{\rm B}\simeq0.242\). Around \(Jt=10\), the values are about \(0.0175\) and \(0.390\), and around \(Jt=15\) they are about \(0.146\) and \(0.713\).

The decomposition shows why. During much of the quantum evolution, \(G_R^{\rm Q}\) and \(\Aenergy^{\rm Q}\) grow together. More information becomes hidden from the spatial record, but the mean energy still resolves a large part of it. In the classical evolution, \(\Delta\Aenergy^{\rm cl}\) becomes negative over much of the stroke, so the energy loses resolving power relative to its initial value. The endpoint process inequality \(\Sigma_{BC}^{\rm Q}<\Sigma_{BC}^{\rm cl}\) therefore comes from a difference that develops during the dynamics, not from the final maximum-entropy calculation alone.

\suppnote{Numerical checks and error budget}
\label{supp:note-error-budget}

The calculations use deterministic quantum propagation and constrained optimization together with classical Hamiltonian integration and phase-space quadrature. We check the main numerical errors separately: quantum propagation, maximum-entropy reconstruction, classical time stepping, and classical quadrature or sampling. The main checks are summarized here and shown in Figs.~\ref{fig:supp-matched-validation} and \ref{fig:supp-refinement} where useful.

For the quantum particle-number scan, the continuous-ramp propagation is performed with the adaptive DOP853 solver described in Supplementary Note~\ref{supp:note-quantum-numerics}. Across \(Q=2,\ldots,7\), the numerical drift of the fine von Neumann entropy remains of order \(10^{-8}\) or smaller. The subsequent energy-and-record maximum-entropy reconstruction satisfies all target constraints with a maximum absolute residual no larger than \(7.4\times10^{-6}\), including at \(Q=7\), where the quantum--classical irreversibility gap is smallest. These deterministic numerical errors are negligible on the scale of the classical sampling uncertainty and of the reported finite-\(Q\) differences.

The strict matched-preparation calculation provides an independent deterministic convergence test for the reduced classical thermodynamic integrals. For the initial fine entropy, Gauss--Legendre order \(12\) gives \(S_B^{\rm cl}/k_{\rm B}=-6.3805447218\), while orders \(16\) through \(40\) give
\begin{equation}
\frac{
S_B^{\rm cl}
}{
k_{\rm B}
}
=
-6.3805447275
\end{equation}
to the displayed precision. For the fixed representative endpoint constraints used specifically in the quadrature-convergence test, the corresponding energy-and-record entropy changes from \(S_{E,R}^{\rm cl}(C^-)/k_{\rm B}=-4.4480335456\) at order \(12\) to
\begin{equation}
\frac{
S_{E,R}^{\rm cl}(C^-)
}{
k_{\rm B}
}
=
-4.4480335483
\end{equation}
from order \(16\) onward. These values refer to the fixed convergence-test endpoint and should not be confused with the final production estimate \(S_{E,R}^{\rm cl}(C^-)/k_{\rm B}=-4.4480560\pm0.0022347\), for which the endpoint constraints themselves are obtained from the target-adapted trajectory ensemble. The difference between these two endpoint numbers therefore reflects the sampled endpoint inputs, not unresolved deterministic quadrature error.

Sampling convergence of the strict preparation is shown in Fig.~\figpanel{fig:supp-matched-validation}{a}. Using \(2^{14}\) trajectories per macro-bin gives
\begin{equation}
\frac{
\Sigma_{BC}^{\rm cl,matched}
}{
k_{\rm B}
}
=
1.93644
\pm
0.00295,
\end{equation}
whereas the production calculation with \(2^{16}\) trajectories per macro-bin gives
\begin{equation}
\frac{
\Sigma_{BC}^{\rm cl,matched}
}{
k_{\rm B}
}
=
1.93249
\pm
0.00223.
\end{equation}
The two estimates differ by approximately \(1.1\) combined standard errors and remain far above the quantum value \(\Sigma_{BC}^{\rm Q}/k_{\rm B}=0.641610\). Figure~\figpanel{fig:supp-matched-validation}{b} independently confirms that the deterministic thermodynamic integrations are already converged well beyond this sampling precision.

For the thermal protocol-duration scan, halving the classical Strang-splitting step from \(J\Delta t=0.05\) to \(0.025\) changes \(\Sigma_{BC}^{\rm cl}\) by at most \(0.109\%\) among the tested refinement points, while increasing the Sobol resolution from \(2^{15}\) to \(2^{16}\) changes it by at most \(0.322\%\); see Figs.~\figpanel{fig:supp-refinement}{a} and \figpanel{fig:supp-refinement}{b}. For the larger \(L=6\) calculations, the corresponding Sobol-resolution changes are \(0.428\%\) and \(0.365\%\) in magnitude for \(Q=3\) and \(Q=4\), with a further nested \(2^{17}\) check producing only approximately \(0.05\%\) additional paired shifts. Halving the propagation step changes the same \(L=6\) results by approximately \(0.02\%\) [Figs.~\figpanel{fig:supp-refinement}{c} and \figpanel{fig:supp-refinement}{d}].

The equilibrium parts of the cycle closure do not introduce a new sampled entropy uncertainty. The \(D\rightarrow A\) return is an exact Hamiltonian scaling for which the equilibrium state is unchanged, so \(\Sigma_{DA}=0\) analytically. At the cold-side interface, \(H_{C^-,R}\) and the switch work are algebraic functions of the fitted endpoint constraints, while the fixed-Hamiltonian \(C^+\rightarrow\bar C\) relaxation has zero net heat by the exact energy identity derived in Note~\ref{supp:note-cycle} of the SM \cite{SuppMat}. The quasistatic \(\bar C\rightarrow D\) heat is then obtained from the explicit equilibrium thermodynamic path. A dedicated representative-point quantum audit verifies the \((E,R)\) reconstruction to \(1.6\times10^{-10}\), the \(H_{C^-,R}\) energy identity and direct-work closure to \(1.7\times10^{-10}J\), and the reservoir entropy balance to \(5.8\times10^{-15}k_{\rm B}\). A fresh classical QMC closure audit verifies the corresponding energy, work, and entropy closure relations to better than \(2.7\times10^{-15}\) for its realized endpoint; see Note~\ref{supp:cycle-audit} of the SM \cite{SuppMat}. Because that classical audit is a separate stochastic rerun, it is not substituted for the production ensemble means. The retained cycle-entropy uncertainty is therefore the uncertainty of the \(B\rightarrow C^-\) production calculation and its endpoint reconstruction. For the representative four-particle comparison, the same-\(Q_h\) matching uses \(\kappa_c^{\rm Q}=2.7\) and \(\kappa_c^{\rm cl}=2.564983\), giving the common target \(Q_h=5.242261J\) at the ensemble-mean level. The resulting efficiency-gap uncertainty follows directly from the classical uncertainty in \(\Sigma_{BC}\).

These independent checks address different numerical approximations and lead to the same conclusion: the reported quantum--classical ordering is resolved well beyond the deterministic integration errors and is stable under the tested classical time-step and sampling refinements. No extrapolation from an unconverged estimator is required for the sign or characteristic scale of the finite-particle reduction.

\suppnote{Summary of the comparison and checks}
\label{supp:note-controls-summary}

The comparison is arranged so that the final quantum--classical difference cannot be blamed on a simple mismatch of models, records, initial data, or cycle resources.

First, the microscopic energy structure is matched. The finite-particle Bose--Hubbard model and the number-conserving classical field have the same coherent-state energy landscape and are driven by the same confinement protocol. Their dynamics are still different: the quantum state explores the full fixed-particle Hilbert space, while the classical field follows Hamiltonian flow.

Second, both systems are viewed through the same coarse right-half record. The record-only representative is \(\bar\rho_R=\cG_R[\rho]\), and the exact state is \(\rho=\bar\rho_R+\chi_R\), as in Ref.~\cite{ahmadi2026endogenous}. Adding the mean energy gives the second representative \(\bar\rho_{E,R}\) and the remaining operator difference \(\chi_{E,R}=\rho-\bar\rho_{E,R}\). The exact relation
\begin{equation}
\chi_R
=
\left(
\bar\rho_{E,R}-\bar\rho_R
\right)
+
\chi_{E,R}
\end{equation}
separates what the mean energy resolves from what remains hidden even after both \(E\) and \(R\) are kept. The classical construction has the same form.

The corresponding state-level entropy identity is
\begin{equation}
G_R
=
\Aenergy+\Ires.
\end{equation}
Here \(\Ires\) is the residual information gap of a state, not an entropy generation. We reserve \(\Sigma\) for process or cycle entropy generation. For the isolated process \(B\rightarrow C^-\), the fine entropy is conserved and the canonical initial state has \(\Ires(B)=0\), so the process entropy generation satisfies \(\Sigma_{BC}=\Ires(C^-)\). This equality is special to the present preparation and process; the two symbols still describe different kinds of quantities.

The mechanism is then easy to read. The quantum spatial record can relax more strongly, so the lower quantum entropy generation is not caused by weaker macroscopic relaxation. The difference comes from the mean energy: after the finite-particle quantum evolution it removes more of the ambiguity left by the spatial record. The strict-preparation test reaches the same conclusion even when the initial classical record and mean energy are forced to match the quantum values.

The cycle turns this state gap into a physical entropy cost. At \(C^-\), the controller uses only the measured mean energy and record probabilities to build \(\bar\rho_{E,R}(C^-)\) and the controlled Hamiltonian \(H_{C^-,R}\). The work-only switch to this Hamiltonian leaves the microscopic state unchanged. At \(C^+\), the cold reservoir is attached once and remains attached until \(D\). The actual state first relaxes to \(\bar\rho_{E,R}(C^-)\) at fixed Hamiltonian with zero net heat. During this relaxation, the residual information gap closes and the fine entropy increases by the same amount \(\Sigma_{BC}\). The same reservoir then remains attached while the Hamiltonian is changed quasistatically to \(H_D\). The final isolated scaling returns the system to \(A\).

The step-by-step retained-description balance and the physical entropy balance therefore give the same cycle entropy generation, \(\Sigma_{\rm cyc}=\Sigma_{BC}\). This is not double counting. The isolated process creates the gap relative to the retained description; the later cold relaxation physically removes that same hidden distinction. The switch work is still part of the ordinary energy balance and is included in the independent work audit.

The cycle does not require full tomography. After \(C^-\), all controls are fixed by the retained mean energy, the record probabilities, the reservoir temperatures, and known model parameters. The assumed quantum controls include an overall scale of \(H_0\) and a record-dependent potential \(F_{C^-}(N_R)=\sum_r\alpha_{r,C^-}P_r\), with the matching classical term.

Finally, every change of expansion duration creates a new endpoint, so the energy-and-record representative and the controlled Hamiltonian are rebuilt for each duration. The additional checks show that the lower quantum process cost is not tied to one lattice size, one interaction strength, or one sharp classical partition.

All of these tests support the same conclusion. Both microscopic descriptions preserve fine information during the isolated dynamics, but the chosen thermodynamic record leaves a smaller residual information gap after the finite-particle quantum evolution. The resulting process and cycle entropy costs are therefore smaller on the quantum side, and the difference becomes weaker as the classical-field regime is approached.

\end{document}